\documentclass[usenatbib]{mn2e}
\usepackage{graphicx,tabularx,amsmath,amssymb,upgreek,wasysym,mathtools,pdflscape,longtable,overpic}
\numberwithin{equation}{section}
\title[Simulating star formation in Ophiuchus]{Simulating star formation in Ophiuchus}
\author[O. Lomax, A. P. Whitworth, D. A. Hubber, D. Stamatellos and S. Walch]{O. Lomax\thanks{E-mail: oliver.lomax@astro.cf.ac.uk}$^1$, A. P. Whitworth$^1$, D. A. Hubber$^{2,3}$, D. Stamatellos$^4$ and S. Walch$^{5,6}$\\
$^1$School of Physics and Astronomy, Cardiff University, Cardiff CF24 3AA, UK\\
$^2$University Observatory, Ludwig-Maximilians-University Munich, Scheinerstr.1, D-81679 Munich, Germany\\
$^3$Excellence Cluster Universe, Boltzmannstr. 2, D-85748 Garching, Germany\\
$^4$Jeremiah Horrocks Institute, University of Central Lancashire, Preston, Lancashire, PR1 2HE, UK\\
$^5$Physikalisches Institut der Universit{\"a}t zu K{\"o}ln, Z{\"u}lpicher Strasse 77, D-50937 Cologne, Germany\\
$^6$Max-Planck-Institut f{\"u}r Astrophysik, Karl-Schwarzschild-Str. 1, 85748 Garching}

\begin{document}
\pagerange{\pageref{firstpage}--\pageref{lastpage}} \pubyear{2013}
\maketitle
\label{firstpage}

\begin{abstract}
We have simulated star formation in prestellar cores, using SPH and initial conditions informed by observations of the cores in Ophiuchus. Because the observations are limited to two spatial dimensions plus radial velocity, we cannot infer initial conditions for the collapse of a particular core. However, with a minimum of assumptions (isotropic turbulence with a power-law spectrum, a thermal mix of compressive and solenoidal modes, a critical Bonnor-Ebert density profile) we can generate initial conditions that match, in a statistical sense, the distributions of mass, projected size and aspect ratio, thermal and non-thermal one-dimensional velocity dispersion, observed in Ophiuchus. The time between core-core collisions in Ophiuchus is sufficiently long, that we can simulate single cores evolving is isolation, and therefore we are able to resolve masses well below the opacity limit. We generate an ensemble of 100 cores, and evolve them with no radiative feedback from the stars formed, then with continuous radiative feedback, and finally with episodic radiative feedback. With no feedback the simulations produce too many brown dwarfs, and with continuous feedback too few. With episodic radiative feedback, both the peak of the protostellar mass function (at $\sim 0.2\,{\rm M}_{_\odot}$) and the ratio of H-burning stars to brown dwarfs are consistent with observations. The mass of a star is not strongly related to the mass of the core in which it forms. Low-mass cores ($M_\textsc{core}\sim0.1\,\mathrm{M}_{_\odot}$) tend to collapse into single objects, whereas high-mass cores ($M_\textsc{core}\gtrsim1\,\mathrm{M}_{_\odot}$) usually fragment into several objects with a broad mass range.
\end{abstract}


\vspace{0.5cm} 

\section{Introduction}%

The origin of the stellar initial mass function (IMF), and why it appears to be approximately universal, remain major problems in star formation. Two main scenarios have been proposed to addresses these problems.

In {\it Competitive Accretion}, part of a molecular cloud collapses, thereby providing a gas-rich environment in which stars form. The most massive stars preferentially accrete the most gas; the least massive stars may be ejected from the system by $N$-body interactions and cease accreting \citep[e.g.][]{BBCP97,RC01,BVB04,BB05,B09c}. This scenario naturally yields a distribution of stellar masses similar in shape to the IMF \citep{K01,C05}, with a peak close to the cloud's original Jeans mass.

In {\it Turbulent Fragmentation}, molecular clouds do not collapse on a global scale, but turbulent flows within the cloud produce dense cores \citep[e.g.][]{PN02,HC08,HC09}, roughly between $0.01\,\mathrm{pc}$ and $0.1\,\mathrm{pc}$ across. Cores are usually well separated from one another, and those that are gravitationally bound collapse to form stars \citep[e.g.][]{AWB93,AWB00}. Star formation in cores has been simulated numerically by \citet{B98,B00,HBB01,MH03, GW04,DCB04a,DCB04b,GWW04,GWW06,WBWNG09,WNWB10,WWG12}.

Recent millimetre and submillimetre observations of star forming regions \citep[e.g.][]{MAN98,TS98,JWM00,MAWB01,JFMM01,SSGK06,EYG06,JB06,NW-T07,ALL07,EES08,SNW-T08,RLM09,KAM10} have shown that the core mass function (CMF), is similar in shape to the IMF, but shifted upwards in mass by a factor of 3 to 5. From this it is commonly inferred that the shape of the IMF is directly inherited from the CMF, i.e. each core collapses and converts roughly the same fraction of its mass into the same number of protostars. \citet{HWGW13} have shown that this is only compatible with the observed stellar multiplicity statistics if a typical core spawns 4 or 5 stars.

The L1688 cloud (hereafter Ophiuchus) is a region where stars are forming in cores. Observations of Ophiuchus by \citet{MAN98} and \citet{ABMP07} (hereafter MAN98 and ABMP07) show that the cores are well separated and unlikely to interact dynamically with one another before they collapse. In this paper we use SPH to simulate star formation in these cores, using initial conditions informed as closely as possible by the MAN98 and ABMP07 observations. A limitation of our approach is that cores tend to be embedded in larger scale clumps and filaments, and material that would accrete onto the cores as they collapse is not included in the simulations.

Deriving reliable initial conditions for simulations of core collapse is a difficult inverse problem. Core masses can be estimated in a relatively straightforward way from observed submillimetre continuum fluxes. However, the intrinsic three-dimensional sizes and shapes of cores can only be constrained by fitting the parameters of a credible model to observed data \citep[see][hereafter LWC13]{LWC13}. The detailed three-dimensional motions within cores are even less well constrained by observations of radial velocity profiles. A key element of this paper is to generate an ensemble of cores which, in a statistical sense, resemble those observed in Ophiuchus.

In Section \ref{observations} we review the data used to constrain the distribution of core masses, sizes and velocity dispersions in Ophiuchus. We use these data to calibrate a multivariate lognormal distribution from which we can draw representative core masses, sizes and velocity dispersions. In Section \ref{initial_conditions} we describe how we set up the initial conditions for a core; this includes assigning an ellipsoidal shape to the core and giving it a velocity field and a density profile. In Section \ref{numerics} we outline the SPH code used. For each core, we perform simulations with no radiative feedback, continuous radiative feedback, and episodic radiative feedback. In Section \ref{results} we present the results of the simulations and in Section \ref{conclusions} we discuss and summarise the results.

\section{Observations}\label{observations}%

Ophiuchus is a star forming region $\sim\!140\,\mathrm{pc}$ away from the Sun \citep{M08}. It has been well observed in the millimetre and submillimetre continuum \citep[e.g.][]{MAN98,SSGK06,SNW-T08}, which traces the column density of dust, and allows observers to estimate the masses and sizes of cores. \citet{L99} calculates that Ophiuchus has total mass $M_\textsc{oph}\!\sim\!1\,{\rm to}\,2\times\!¬10^3\mathrm{M}_{_\odot}$ and diameter $D_\textsc{oph}\!\sim\!1\mathrm{pc}$. The cores in Ophiuchus are grouped into clumps\footnotemark, labelled Oph-A, Oph-B1, Oph-B2, Oph-C, Oph-D, Oph-E and Oph-F.

\footnotetext{MAN98 refer to these groups as ``cores'', and to the smaller denser condensations within them as ``clumps''. We have adopted the now near-universal opposite convention that the small dense condensations are cores, and they are grouped in clumps.}

We base the initial conditions for our simulations on the observations of MAN98 and ABMP07. MAN98 present continuum observations of cores with a completeness limit of $\sim\!0.1\,\mathrm{M}_{\odot}$. These data are complemented by observations of core velocity dispersions in Ophiuchus by ABMP07. There are more recent measurements of core masses and sizes in Ophiuchus \citep[e.g.][]{SSGK06,SNW-T08}, but they contain fewer cores with an associated ABMP07 velocity dispersion than the MAN98 data. Further details of the MAN98 and ABMP07 observations are given in the next two sections.

\subsection{MAN98}%
\label{man98sec}

MAN98 map dust emission at $1.3\,\mathrm{mm}$ using the MPIfR bolometer array on the IRAM $30\,\mathrm{m}$ telescope. The beam size is $11''$, corresponding to $1500\,{\rm AU}$ at Ophiuchus, and they are sensitive to hydrogen column-densities $N\!>\!10^{22}\,{\rm H}\,{\rm cm}^{-2}$. Their map covers an area of $\sim\!480\,\mathrm{arcmin}^2$ and includes the Oph-A, Oph-B1, Oph-B2, Oph-C, Oph-D, Oph-E and Oph-F clumps. MAN98 use a multi-scale wavelet analysis to extract 61 cores, 36 of which are resolved. Two dimensional Gaussian distributions are fitted to the cores and the core dimensions are given by the FWHM of the major and minor axes. The cores have masses between $\sim\!0.1\,{\rm M}_{_\odot}$ and $\sim\!3\,{\rm M}_{_\odot}$, and dimensions between $\sim\!1000\,{\rm AU}$ and $\sim\!20,000\,{\rm AU}$.

The core masses and sizes used in this paper are slightly different from those given in MAN98. The mass of a prestellar core is calculated using
\begin{equation}
   \label{m_estimate}
   M_\textsc{core}=\frac{S(\lambda)D^2}{\kappa(\lambda)B(\lambda,T)}\,,
\end{equation}
where $S(\lambda)$ is the monochromatic flux from the core at wavelength $\lambda$, $D$ is the distance to the core, $\kappa(\lambda)$ is the mass opacity, and $B(\lambda,T)$ is the Planck function at temperature $T$. For $\lambda=1.3\,\mathrm{mm}$, MAN98 adopt $\kappa(1.3\,\mathrm{mm})=0.005\,\mathrm{cm^2\,g^{-1}}$. The dust temperature estimates used by MAN98 have been revised by \citet{SWW07}. These temperatures, $T_\textsc{man98}$ and $T_\textsc{sww07}$ are given in Table \ref{temperatures}. The distance to Ophiuchus used by MAN98, $D_\textsc{man98}=160\,\mathrm{pc}$, has also been revised, by \citet{M08}, to $D_\textsc{m08}=139\pm6\,\mathrm{pc}$. To account for this, each mass given by MAN98 must be transformed thus:
\begin{equation}
   \label{m_conversion}
   M_\textsc{man98}\to\frac{B(1.3\,\mathrm{mm},T_\textsc{man98})D_\textsc{m08}^2}{B(1.3\,\mathrm{mm},T_\textsc{sww07})D_\textsc{man98}^2}M_\textsc{man98}\,,
\end{equation}
Similarly, each size must be transformed thus:
\begin{equation}
   R_\textsc{man98}\to\frac{D_\textsc{m08}}{D_\textsc{man98}} R_\textsc{man98}\,.
\end{equation}
The dust temperatures $T_\textsc{sww07}$ are cooler than $T_\textsc{man98}$, which increases the estimated core masses. However, the revised distance $D_\textsc{m08}$ is less than $D_\textsc{man98}$, which decreases the core masses. The combined effect modestly increases the masses of cores in Oph-A, Oph-E and Oph-F, whilst leaving the masses of cores in Oph-B, Oph-C and Oph-D essentially unchanged (see Table \ref{temperatures}).

After these adjustments have been made, the MAN98 CMF peaks at $M_{_{\rm PEAK}}\!\sim\!0.3\,{\rm M}_{_\odot}$. This is lower than the values found for the Ophiuchus cores by other groups, i.e. $M_{_{\rm PEAK}}\!\sim\!0.9\,{\rm M}_{_\odot}$ \citep{SSGK06} and $M_{_{\rm PEAK}}\!\sim\!0.4\,{\rm M}_{_\odot}$ \citep{SNW-T08}. The differences in $M_{_{\rm PEAK}}$ must be attributed to a combination of the different wavelengths used; the different assumptions made about dust opacities, dust temperatures, and source distances; and the different procedures used to identify cores, distinguish them from the background, and correct for incompleteness at low masses and poor statistics at high masses. We also note that studies of other star formation regions, e.g. Perseus \citep{EES08}, the Pipe Nebula \citep{RLM09} and Aquilla \citep{KAM10} find CMFs with $M_{_{\rm PEAK}}$ between $\sim\!0.6\,{\rm M}_{_\odot}$ and $\sim\!1.0\,{\rm M}_{_\odot}$. We have therefore doubled the MAN98 masses, so that the peak of the CMF is at $M_{_{\rm PEAK}}\sim\!0.6\,{\rm M}_{_\odot}$, i.e. between the \citet{SSGK06} and \citet{SNW-T08} values, and within the range found for other star formation regions.

\begin{table}
   \centering
   \begin{tabular}{cccc}\hline
      Region & $T_\textsc{man98}$ & $T_\textsc{sww07}$ & $\times M_\textsc{man98}$ \\
      & (K) & (K) & \\\hline
      Oph-A & 20 & 11 & 1.77 \\
      Oph-B & 12 & 10 & 1.01 \\
      Oph-C & 12 & 10 & 1.01 \\
      Oph-D & 12 & 10 & 1.01 \\
      Oph-E & 15 & 10 & 1.40 \\
      Oph-F & 15 & 10 & 1.40 \\
      Oph-J & - & 10 & - \\\hline
   \end{tabular}
   \caption{Dust temperatures from MAN98 and SWW07. The fourth column gives the correction factor applied to the MAN98 core masses when we adopt $T=T_\textsc{sww07}$ and $D=139\,\mathrm{pc}$.}
   \label{temperatures}
\end{table}

\subsection{ABMP07}%

ABMP07 measure the width of the $\mathrm{N_2H^+}$(1--0) emission line from 27 prestellar cores in Ophiuchus. Of these 27 cores, 26 have mass estimates from MAN98. The line width gives the radial velocity dispersion, $\sigma_\textsc{1d}$, of the $\mathrm{N_2H^+}$ molecules in the  core. In general there are thermal and non-thermal contributions to the line width, $\sigma_\textsc{1d}^2=\sigma_\textsc{t}^2+\sigma_\textsc{nt}^2$. The thermal contribution is given by
\begin{equation}
   \sigma_\textsc{t}^2=\frac{k_\textsc{b}T}{m_\textsc{mol}}\,,
\end{equation}
where $m_\textsc{mol}=4.8\times10^{-23}\,\mathrm{g}$ is the mass of an $\mathrm{N_2H^+}$ molecule. The non-thermal contribution, 
\begin{equation}
    \sigma_\textsc{nt}=\left(\sigma_\textsc{1d}^2-\frac{k_\textsc{b}T_\textsc{sww07}}{m_\textsc{mol}}\right)^{1/2}\,,
\end{equation}
gives the magnitude of macroscopic gas motions, i.e. turbulence, rotation and contraction or expansion. Since we are adopting the reduced temperatures from SWW07, we derive somewhat higher values of $\sigma_\textsc{nt}$ than ABMP07, between $\sim\!0.05\,{\rm km\,s^{-1}}$ and $\sim\!0.3\,{\rm km\,s^{-1}}$.

ABMP07 also measure the bulk radial velocities of cores in Ophiuchus. From the dispersion in the bulk radial velocities of the cores within a clump, they estimate that, on average, a core should collide with another core after a few times $10^5\,\mathrm{yrs}$. In contrast, the individual cores  (see Section \ref{lndist} below, and Table \ref{onlinematerial} in online material) have freefall times $t_{_{\rm FF}}\sim 4^{+4}_{-2}\times 10^4\,\mathrm{yrs}$. Thus, whereas interactions between cores may have been important in delivering the  distribution of core properties we observe today, the future evolution of these cores should not be much influenced by interactions. This conclusion is born out in our simulations, where the first protostar usually forms after about one freefall time, and any subsequent star formation tends to be concentrated in the vicinity of this protostar. As a further precaution, we terminate the simulations after $2\times 10^5\,\mathrm{yrs}$.

Treating individual cores in isolation allows us to perform simulations with sufficiently high resolution to capture the formation of protostars close to the opacity limit, and to perform many realisations, but it deprives us of the possibility of considering the effects of -- for example -- accretion flows and tidal perturbations from the surroundings. A large-scale simulation treating the whole of Ophiuchus with the same resolution is currently beyond our computational resources, but it should be feasible to simulate a whole clump (e.g. Oph-A).

\subsection{Lognormal distribution}%
\label{lndist}

\begin{figure*}
   \centering
   \includegraphics[width=0.7\textwidth]{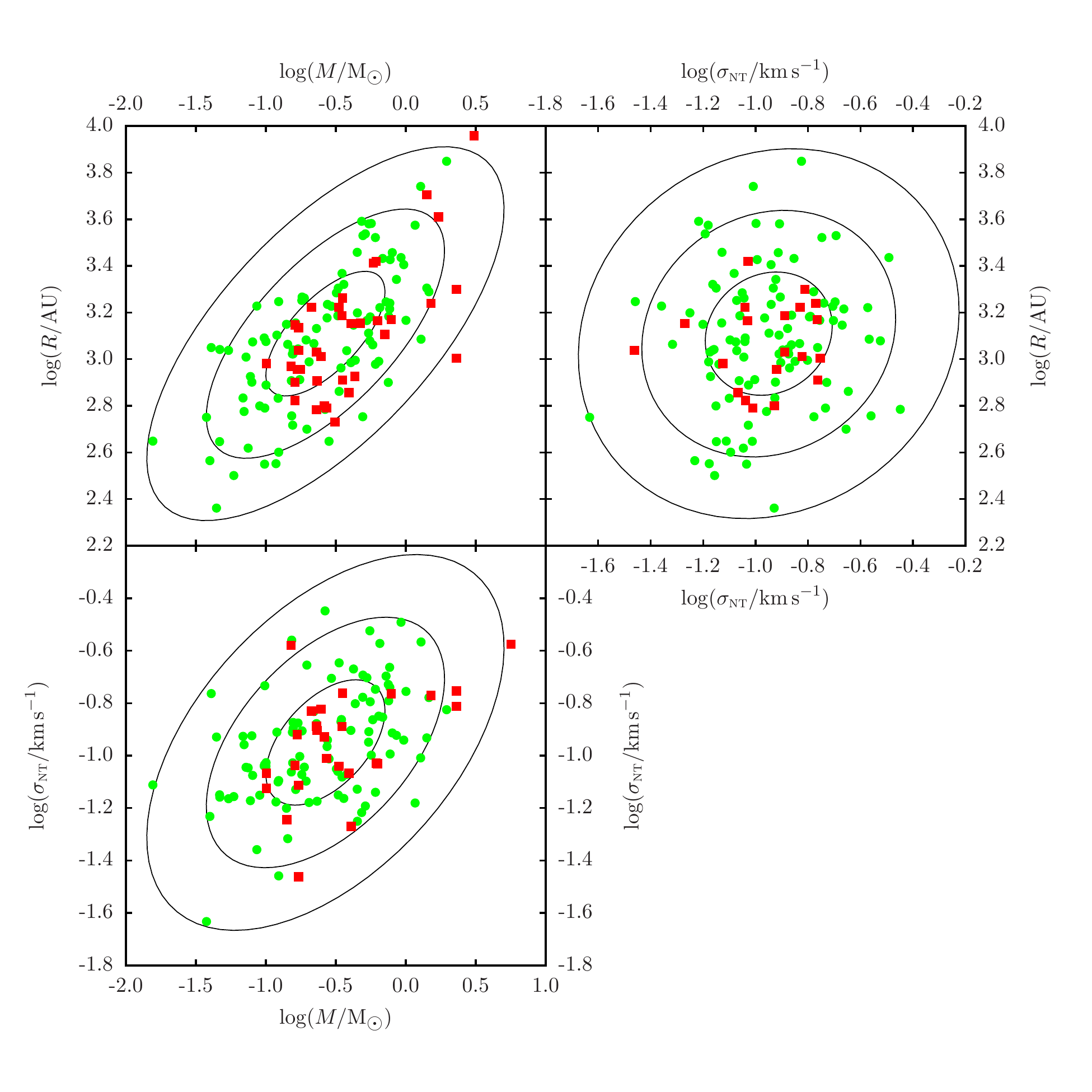}
   \caption[Lognormal distribution of core parameters.]{The multivariate lognormal distribution, $P(\boldsymbol{x})$ where \mbox{$\boldsymbol{x}=(\log(M),\log(R),\log(\sigma_\textsc{nt}))$}. (a) shows the projections through $\log(\sigma_\textsc{nt})$; (b) through $\log(M)$; and (c) through $\log(R)$. The concentric ellipses show the $1\sigma$, $2\sigma$ and $3\sigma$ regions of the distribution. The green circles are randomly drawn points from $P(\boldsymbol{x})$. The red squares are the observational data from MAN98 and ABMP07. Because there is a limted number of cases for which more than one of these parameters is known, the distribution of obervational points is shifted relative to the fitted distribution. }
   \label{core_dist}
\end{figure*}

Together, MAN98 and ABMP07 provide 61 core masses, 36 mean core radii and 27 core velocity dispersions; 20 cores have all three. We use these data to calibrate a multivariate lognormal distribution of core properties. To within the errors of small number statistics, and modulo the assumption of a lognormal distribution, this reproduces statistically the observed properties of cores in Ophiuchus, \emph{and} their correlations. Specifically, we draw values of $\boldsymbol{x}\equiv(\log(M),\log(R),\log(\sigma_\textsc{nt}))$ from a probability density
\begin{equation}
   P(\boldsymbol{x})=\frac{1}{(2\uppi)^{3/2}|\boldsymbol{\varSigma}|}\exp\left(-\frac{1}{2}(\boldsymbol{x}-\boldsymbol{\mu})^\mathrm{T}\boldsymbol{\varSigma}^{-1}(\boldsymbol{x}-\boldsymbol{\mu}) \right)\,,
   \label{obsdist}
\end{equation}
where
\begin{equation}
   \boldsymbol{\mu}\equiv
      \begin{pmatrix}
         \mu_{_M} \\
         \mu_{_R} \\
         \mu_{_{c}}
      \end{pmatrix}\,,
\end{equation}
and
\begin{equation}
      \boldsymbol{\varSigma}\equiv
      \begin{pmatrix}
         \sigma_{_M}^2 & \rho_{_{M,R}}\,\sigma_{_M}\sigma_{_R} & \rho_{_{M,\sigma_\textsc{nt}}}\,\sigma_{_M}\sigma_{_{\sigma_\textsc{nt}}} \\
         \rho_{_{M,R}}\,\sigma_{_M}\sigma_{_R} & \sigma_{_R}^2 & \rho_{_{R,\sigma_\textsc{nt}}}\,\sigma_{_R}\sigma_{_{\sigma_\textsc{nt}}} \\
         \rho_{_{M,\sigma_\textsc{nt}}}\,\sigma_{_M}\sigma_{_{\sigma_\textsc{nt}}} & \rho_{_{R,\sigma_\textsc{nt}}}\,\sigma_{_R}\sigma_{_{\sigma_\textsc{nt}}} & \sigma_{_{\sigma_\textsc{nt}}}^2
      \end{pmatrix}\,.
\end{equation}
Here $\mu_x$ is the arithmetic mean of $\log(x)$, $\sigma_x$ is the standard deviation of $\log(x)$ and $\rho_{x,y}$ is the Pearson's correlation coefficient of $\log(x)$ and $\log(y)$. The values of these terms are given in Table \ref{log_params}\,. Fig. \ref{core_dist} shows the probability density function $P(\boldsymbol{x})$ projected through all three of its dimensions. Superimposed are the observational data points and 100 data points drawn randomly from the distribution.

\begin{table}
   \centering
   \begin{tabular}{llc}\hline
   Parameter & & Value \\\hline
   $\mu_{_M}$ & $[\log(M/\mathrm{M_{\odot}})]$ & -0.57 \\
   $\mu_{_R}$ & $[\log(R/\mathrm{AU})]$ & 3.11 \\
   $\mu_{_{\sigma_\textsc{nt}}}$ & $[\log(\sigma_\textsc{nt}/\mathrm{km\,s}^{-1})]$ & -0.95 \\
   $\sigma_{_M}$ & $[\log(M/\mathrm{M_{\odot}})]$ & 0.43 \\
   $\sigma_{_R}$ & $[\log(R/\mathrm{AU})]$ & 0.27 \\
   $\sigma_{_{\sigma_\textsc{nt}}}$ & $[\log(\sigma_\textsc{nt}/\mathrm{km\,s}^{-1})]$ & 0.20 \\
   $\rho_{_{M,R}}$ && 0.61 \\
   $\rho_{_{M,\sigma_\textsc{nt}}}$ && 0.49 \\
   $\rho_{_{R,\sigma_\textsc{nt}}}$ && 0.11 \\\hline
   \end{tabular}
   \caption{Arithmetic means, standard deviations and correlation coefficients of $\log(M)$, $\log(R)$ and $\log(\sigma_\textsc{nt})$ for cores in Ophiuchus}
   \label{log_params}
\end{table}

\section{Initial conditions}\label{initial_conditions}%

While the mass, size and velocity dispersion of a core are relatively easy to quantify from observations, the same cannot be said for the intrinsic three-dimensional shape, density profile and internal velocity field. Here we define the methodology we adopt for assigning these properties.

\subsection{Shapes}%

In projection, the cores in Ophiuchus have irregular, and often elongated, shapes, which show no significant correlation with mass, size or velocity dispersion. LWC13 show that, if one assumes the intrinsic three-dimensional shape of a core is a randomly oriented triaxial ellipsoid, the observed aspect ratios can be fitted with a one parameter model. Specifically, the relative sizes of the principal axes of the ellipsoid can be generated with
\begin{equation}
  \begin{split}
    A&=1\,,\\
    B&=\exp(\tau\mathcal{G}_\textsc{b})\,,\\
    C&=\exp(\tau\mathcal{G}_\textsc{c})\,,
  \end{split}
\label{EQN:M1}
\end{equation}
where $\mathcal{G}_\textsc{b}$ and $\mathcal{G}_\textsc{c}$ are random numbers drawn from a Gaussian distribution with zero mean and unit standard deviation. For Ophiuchus, a good fit is obtained with $\tau\!=\!0.6$, and improvements to the fit obtained by introducing additional parameters do not appear to be justified (LWC13). Once the axes have been generated using Eqn. (\ref{EQN:M1}), they are re-ordered so that $A\geq B\geq C$ and normalised so that $A=1$. Then, given a mean radius $R$ from the distribution in Eqn. (\ref{obsdist}), the core axes are set to
\begin{equation}
   \begin{split}
      A_\textsc{core}=&\frac{R}{(BC)^{\frac{1}{3}}}\,,\\
      B_\textsc{core}=&BA_\textsc{core}\,,\\
      C_\textsc{core}=&CA_\textsc{core}\,.
   \end{split}
\end{equation}

\subsection{Density profile}%

Observations suggest \citep[e.g.][]{ALL01,HWL01,LMR08} that dense cores often approximate to the density profile of a critical Bonnor-Ebert Sphere \citep{B56}, i.e. $\rho=\rho_{_{\rm C}}{\rm e}^{-\psi(\xi)}$, where $\rho_{_{\rm C}}$ is the central density, $\psi$ is the Isothermal Function, $\xi$ is the dimensionless radius, and the boundary is at $\xi_{_{\rm B}}=6.451$. The FWHM of the column-density through a critical Bonnor-Ebert Sphere corresponds to $\xi_{_{\rm FWHM}}=2.424$, so the density at $(x,y,z)$ is given by
\begin{eqnarray}
\xi&=&\xi_{_{\rm FWHM}}\left(\frac{x^2}{A_\textsc{core}^2}+\frac{y^2}{B_\textsc{core}^2}+\frac{z^2}{C_\textsc{core}^2}\right)^{1/2}\,,\\
\rho&=&\frac{M_\textsc{core}\xi_{_{\rm B}}{\rm e}^{-\psi(\xi)}}{4\pi A_\textsc{core}B_\textsc{core}C_\textsc{core}\psi'(\xi_{_{\rm B}})}\,,\hspace{0.5cm}\xi<\xi_{_{\rm B}}\,,
\end{eqnarray}
where $\psi'$ is the first derivative of $\psi$.

Using this density profile implies that the cores are in hydrostatic equilibrium, contained by an external pressure. For the simulated cores is this paper, this is generally not the case. However, the density profile does reproduce the flat central region and power-law envelope that is often observed in cores \citep[e.g.][]{KWA05,RAP13}.

\subsection{Velocity fields}%

Spectroscopic observations of prestellar cores indicate that their velocity fields may contain ordered rotation \citep[e.g.][]{GBF93}, ordered inward and/or outward motion \citep[e.g.][]{KBLN06} and random turbulence \citep[e.g.][]{DNP95}. If the corresponding contributions to the radial velocity dispersion are $\sigma_\textsc{rot}$, $\sigma_\textsc{rad}$ and $\sigma_\textsc{turb}$, the total is given by
\begin{equation}
   \sigma_\textsc{nt}^2=\sigma_\textsc{rot}^2+\sigma_\textsc{rad}^2+\sigma_\textsc{turb}^2\,.
\end{equation}
For the simplest case of a spherically symmetric core rotating about an axis perpendicular to the line of sight, it requires two parameters to specify the relative contributions to $\sigma_\textsc{nt}$ from rotation, isotropic pulsation and isotropic turbulence. An additional parameter is required if the rotation axis is not perpendicular to the line of sight. The problem becomes even more complicated if the core is ellipsoidal, since the pulsations and turbulence are then unlikely to be isotropic; for example, contraction is likely to be fastest along the shortest axis \citep{LMS65}. We have no way of estimating specific values for any of these parameters. To overcome this problem, we generate a random turbulent velocity field, characterised only by the power exponent (we adopt $\alpha\!=\!2$; cf. Burkert \& Bodenheimer 2000) and the ratio of compressive to solenoidal modes (we adopt the thermal ratio, i.e. ${\cal M}=1/2$), but shift the centre of the core so that it is at the centre of the largest mode; the energy in the largest mode is then invested in ordered rotation and compression, whilst the rest is in smaller-scale turbulence.

\subsubsection{Turbulent motions}\label{turbmotion}%

To construct a dimensionless turbulent velocity field, we work in a $(2\pi)^3$ cubic domain, and populate all modes for which the wave-vector, $\boldsymbol{k}$ has integer components, $(k_x,k_y,k_z)$, in the range 0 to 64. For a power spectrum $P_k\propto k^{-\alpha}$ the energy density in $\boldsymbol{k}$-space is $E_k\propto k^{-(\alpha+2)}$. Hence, with $\alpha\!=\!2$, $E_k\propto k^{-4}$ and the energy is strongly concentrated towards low spatial frequencies (i.e. long wavelengths). The amplitude, $\boldsymbol{a}$, and phase, $\boldsymbol{\varphi}$, of each individual mode, $\boldsymbol{k}$, are set by generating a vector $\boldsymbol{\mathcal G}$ of three random numbers drawn from a Gaussian distribution having zero mean and unit variance, and  a vector $\boldsymbol{\mathcal R}$ of three random numbers drawn from a uniform distribution between zero and one. Then we put
\begin{eqnarray}
\boldsymbol{a}(\boldsymbol{k})&=&\sqrt{E(k)}\,\boldsymbol{\mathcal{G}}\,,\\
\boldsymbol{\varphi}(\boldsymbol{k})&=&2\uppi\,\boldsymbol{\mathcal{R}}\,,
\end{eqnarray}
The corresponding contribution to the dimensionless velocity field is given by
\begin{eqnarray}
\boldsymbol{\hat{v}}(\boldsymbol{k})&=&\boldsymbol{a}(\boldsymbol{k})\cos(\boldsymbol{\varphi}(\boldsymbol{k}))+i\,\boldsymbol{a}(\boldsymbol{k})\sin(\boldsymbol{\varphi}(\boldsymbol{k}))\,,\\\label{fourier}
\boldsymbol{v}(\boldsymbol{x})&=&\frac{1}{(2\uppi)^2}\mathrm{Re}\left(\int\boldsymbol{\hat{v}}(\boldsymbol{k})e^{i\boldsymbol{k}\boldsymbol{x}}\,\mathrm{d}^3k\right)\,;
\end{eqnarray}
Eqn. (\ref{fourier}) is evaluated numerically using the fast Fourier transform library FFTW \citep{FFTW05} to produce values of the velocity field on a $128^3$ grid.

\subsubsection{Ordered motions}\label{orderedmotions}%

\begin{figure*}
  \centering 
   \begin{overpic}[width=0.24\textwidth]
      {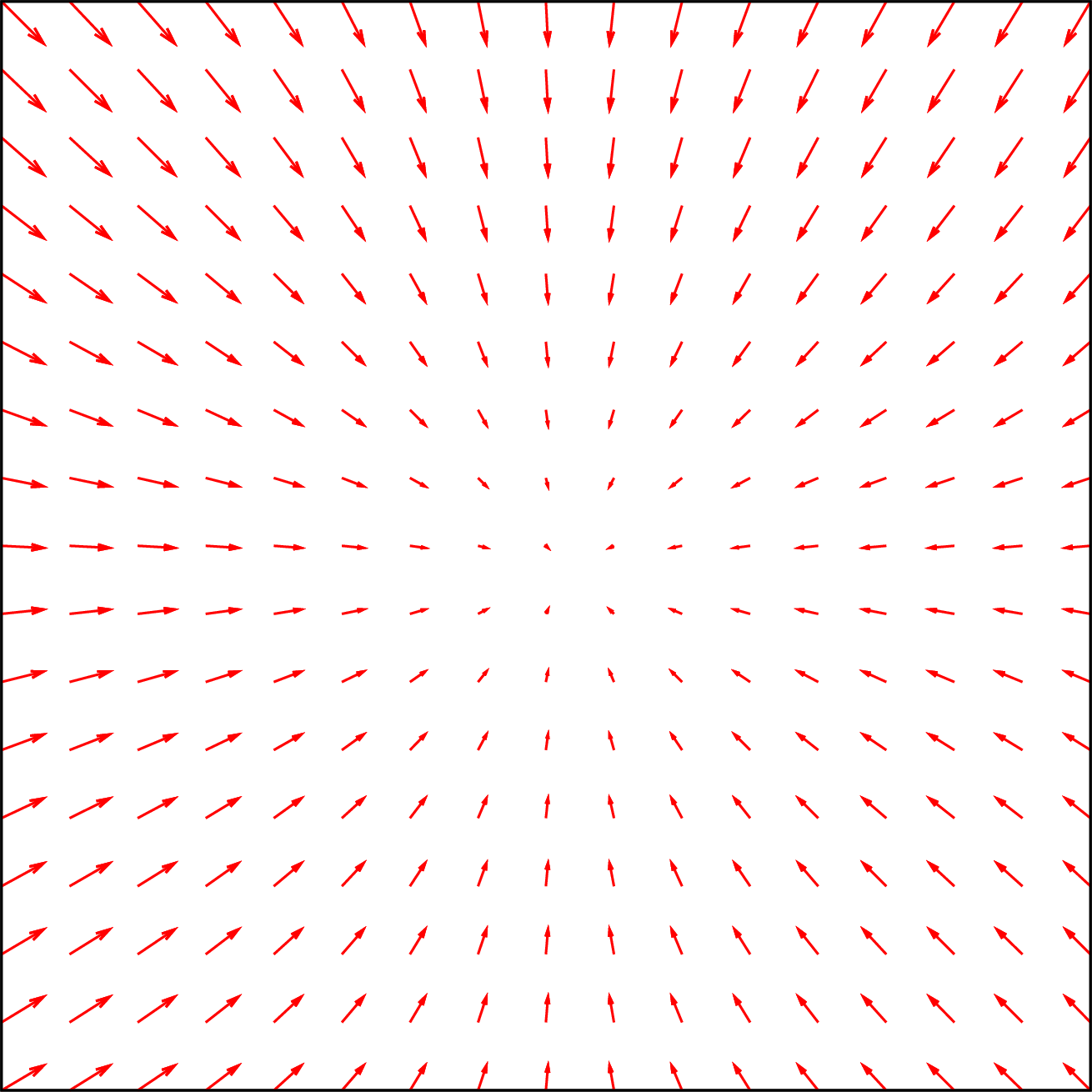}
      \put(10,90){(a)}
   \end{overpic}
      \begin{overpic}[width=0.24\textwidth]
      {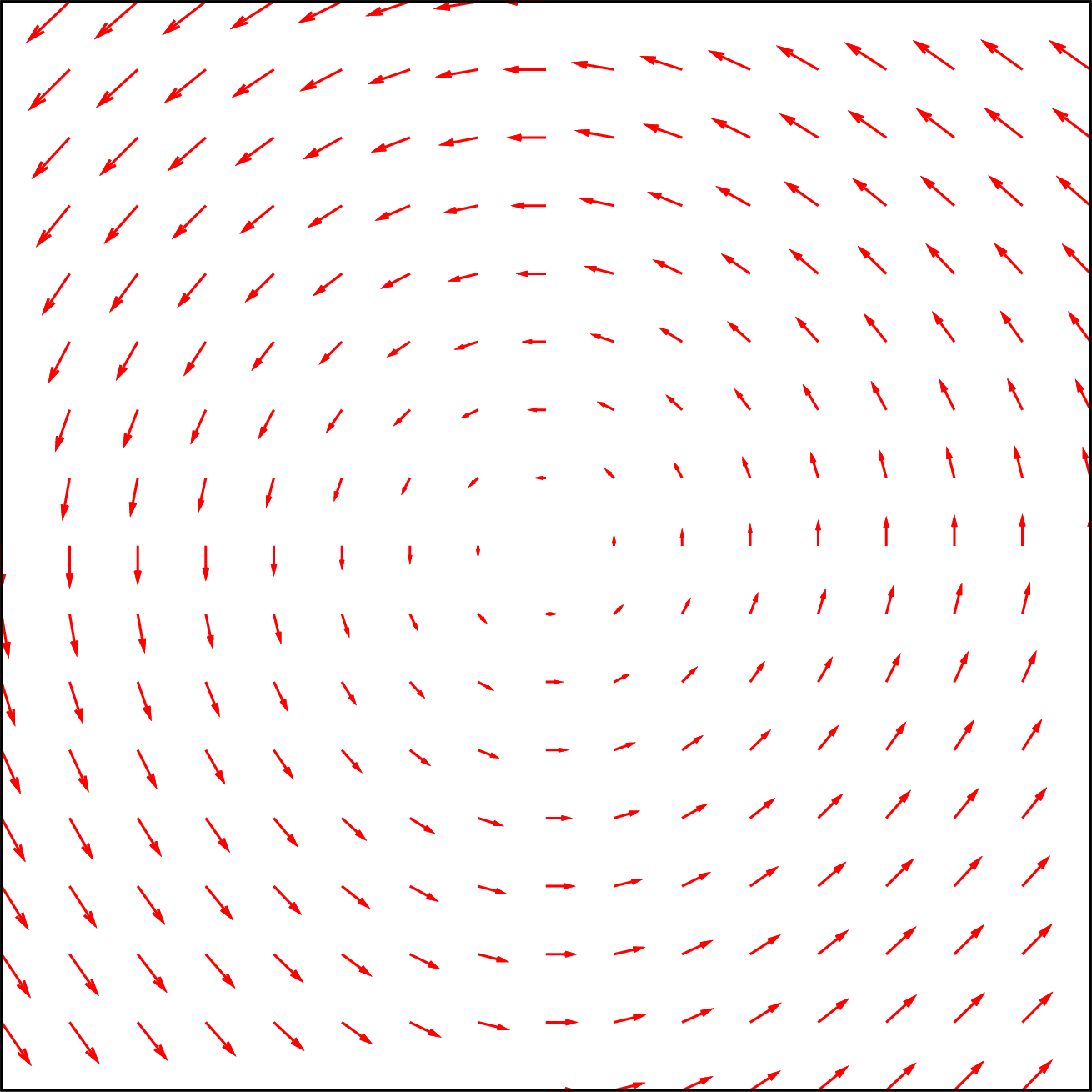}
      \put(10,90){(b)}
   \end{overpic}
      \begin{overpic}[width=0.24\textwidth]
      {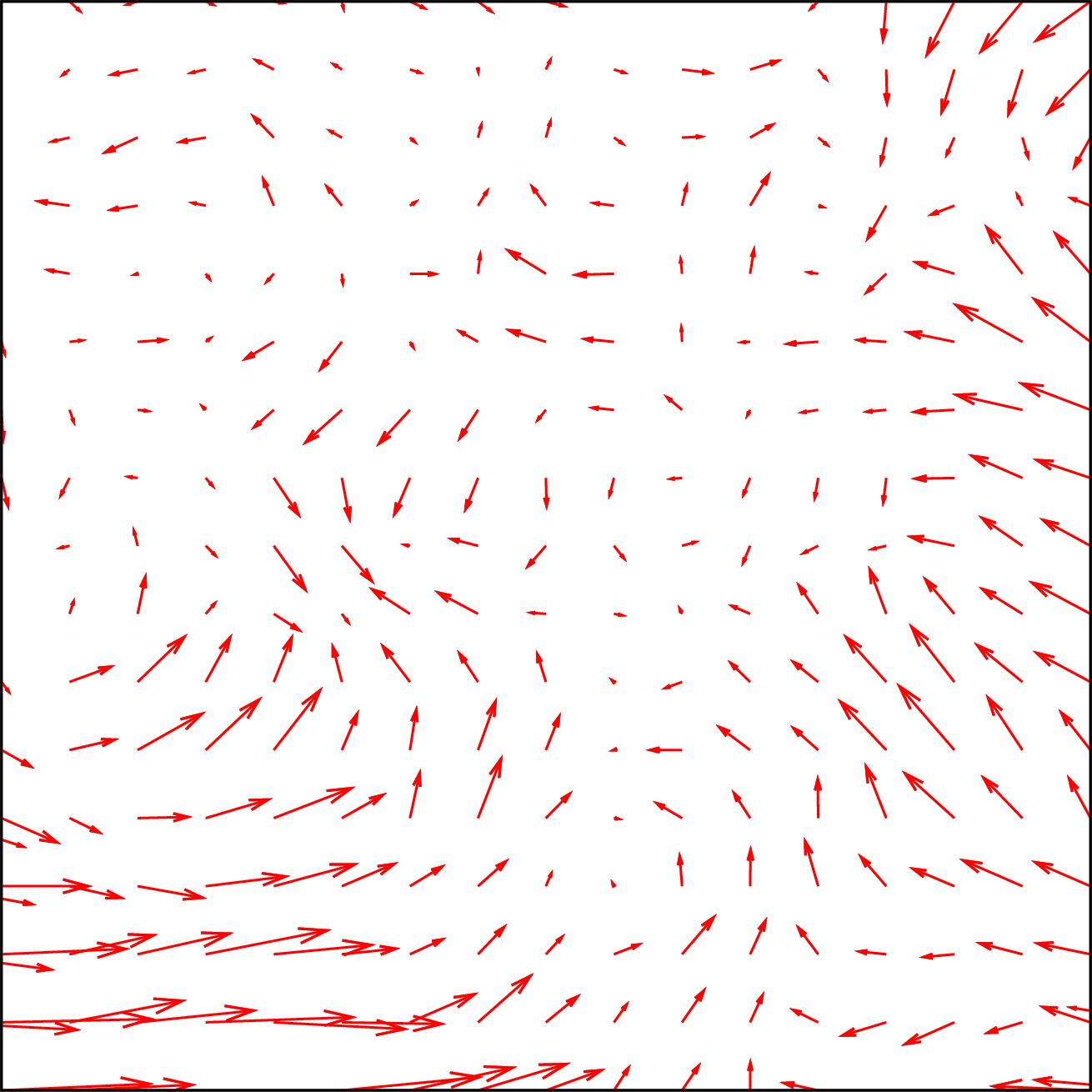}
      \put(10,90){(c)}
   \end{overpic}
      \begin{overpic}[width=0.24\textwidth]
      {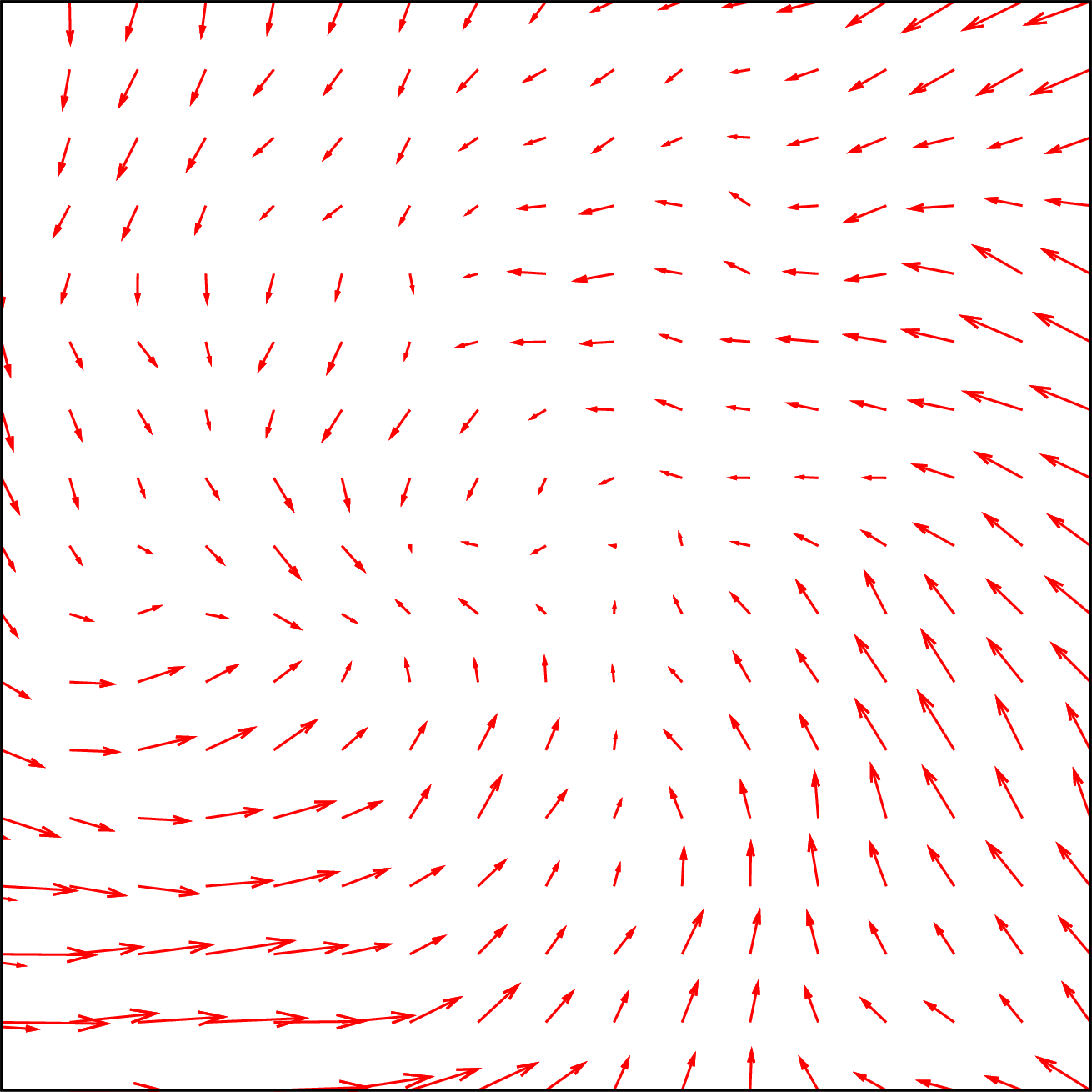}
      \put(10,90){(d)}
   \end{overpic}
  \caption{2D velocity fields generated in an analogous manner to the 3D velocity fields described in Section \ref{turbmotion} and \ref{orderedmotions}. All four frames show the region $-\frac{\uppi}{2}\leq x\leq \frac{\uppi}{2}$ and $-\frac{\uppi}{2}\leq y\leq \frac{\uppi}{2}$. From left to right: (a) a compressive velocity field with $\boldsymbol{a}(1,0)=(-1,0)$, $\boldsymbol{a}(0,1)=(0,-1)$,  $\boldsymbol{\varphi}(1,0)=(\uppi,0)$ and $\boldsymbol{\varphi}(0,1)=(0,\uppi)$; (b) a solenoidal velocity field with $\boldsymbol{a}(1,0)=(0,1)$, $\boldsymbol{a}(0,1)=(-1,0)$,  $\boldsymbol{\varphi}(1,0)=(0,\uppi)$ and $\boldsymbol{\varphi}(0,1)=(\uppi,0)$; (c) a turbulent field with modes generated as described in Section \ref{turbmotion}, but without the $k\!=\!1$ modes; (d) the addition of all three fields giving a superposition of large-scale ordered motion and smaller turbulent modes.}
  \label{vel_fields}
\end{figure*}

We introduce large scale rotation and radial motion into the velocity field by simply modifying the phases and symmetries of the $k\!=\!1$ modes, i.e. $\boldsymbol{k}=(1,0,0),\,(0,1,0)\,{\rm and}\,(0,0,1)$. Specifically, we require that
\begin{equation}
      \begin{split}
         \begin{bmatrix}
            \boldsymbol{a}(1,0,0) \\
            \boldsymbol{a}(0,1,0) \\
            \boldsymbol{a}(0,0,1)
         \end{bmatrix}&=
         \begin{bmatrix}
            r_x & \omega_z & -\omega_y \\
            -\omega_z & r_y & \omega_x \\
            \omega_y & -\omega_x & r_z
         \end{bmatrix}\,,\\
         \begin{bmatrix}
            \boldsymbol{\varphi}(1,0,0) \\
            \boldsymbol{\varphi}(0,1,0) \\
            \boldsymbol{\varphi}(0,0,1)
         \end{bmatrix}&=
         \begin{bmatrix}
            \uppi/2 & \uppi/2 & \uppi/2 \\
            \uppi/2 & \uppi/2 & \uppi/2 \\
            \uppi/2 & \uppi/2 & \uppi/2
         \end{bmatrix}\,.
      \end{split}
\end{equation}
Here $(r_x,r_y,r_z)$ are radial velocity amplitudes and $(\omega_x,\omega_y,\omega_z)$ are rotational velocity amplitudes; the diameter of the core is scaled to $\uppi$, so that it corresponds to the central half wavelength of these modified modes. The outcome of applying this modification is shown in Fig. \ref{vel_fields}\,.

The radial velocity amplitudes are proportional to the rates at which the core expands along the corresponding axis. For example, if $r_x<1$, then the core is contracting along the $x$-axis, and if $r_x>1$, the core is expanding along the $x$-axis. On the assumption that the core is contracting fastest along its shortest axis, and slowest along its longest axis, we re-order the amplitudes so that $r_x\leq r_y\leq r_z$, and orient the core so that its ellipsoidal axes, $A\geq B\geq C$, are respectively aligned with the $x$, $y$ and $z$ axes. The rotational amplitudes define the angular velocity $\boldsymbol{\omega}=(\omega_x,\omega_y,\omega_z)$, giving the axis and rate of core rotation; the axis of rotation has a random direction. The values of $r_x$, $r_y$, $r_z$, $\omega_x$, $\omega_y$ and $\omega_z$ are independently drawn from a Gaussian distribution having zero mean and unit variance. This ensures that the energy in these large-scale components of the velocity field still subscribes to the same $P_k\propto k^{-2}$ power law as the smaller-scale ones that deliver internal turbulence. We stress that only the $k\!=\!1$ modes are modified. All modes with $k>1$ are generated as described in Section \ref{turbmotion}. Hence we have a velocity field composed of ordered rotation, ordered radial motions \emph{and} stochastic turbulent motions, and based on just two parameters, $\alpha$ and ${\cal M}$.

\subsubsection{Interpolation}%

To compute the initial velocity of an SPH particle, $i$, we assign it a position vector,
\begin{equation}
   \boldsymbol{s}_i=\frac{\uppi}{2}\left(\frac{x_i}{A_\textsc{core}},\frac{y_i}{B_\textsc{core}},\frac{z_i}{C_\textsc{core}}\right)\,,
\end{equation}
and estimate its dimensionless velocity by interpolation on the grid of computed values. Finally the dimensionless velocities of the SPH particles are scaled so that their 1D velocity dispersion matches a value of $\sigma_\textsc{nt}$ drawn from Eqn. (\ref{obsdist}).

\section{Numerical Method}\label{numerics}%

\subsection{SPH code}%

We use the \textsc{seren} \citep{HBMW11} implementation of grad-$h$ SPH \citep{SH02,PM04} with smoothing parameter $\eta\!=\!1.2$ (hence typically $\sim 56$ neighbours). Gravity is calculated with a Barnes-Hut tree and a \textsc{gadget}-style multipole acceptance criterion \citep{SH02}. The code invokes multiple particle timesteps and a leapfrog drift-kick-drift integration scheme. Time-dependent artificial viscosity is implemented using the \citet{MM97} formulation with $\alpha=1$, $\alpha_\textsc{min}=0.1$ and $\beta=2\alpha$.

All simulations are set up with $m_\textsc{sph}=10^{-5}\,\mathrm{M}_{\odot}$, so the mass resolution is $\sim 10^{-3}\,{\rm M}_{_\odot}$. A gas concentration that is gravitationally bound and has density higher than $\rho_\textsc{sink}=10^{-9}\,\mathrm{g}\,\mathrm{cm}^{-3}$ is replaced with a sink particle \citep{BBP95} having radius $R_\textsc{sink}\approx0.2\,\mathrm{AU}$. Any SPH particles that subsequently enter this radius and are gravitationally bound to the sink are assimilated by it. We have not used the more sophisticated \textsc{NewSinks} method, presented by \citet{HWW13}, however, for this value of $\rho_\textsc{sink}$, the two methods are reasonably well converged.

Heating and cooling of the gas are treated with the scheme described in \citet{SWBG07}. This takes account of heating by cosmic rays and background radiation, the transport of cooling radiation against dust opacity and molecular opacity, the sublimation of dust, the ionisation and dissociation of hydrogen and the ionisation of helium. It has been extensively tested, and reproduces very accurately the detailed results of \citet{MI00}.

\subsection{Accretion luminosity}%

The dominant contribution to the luminosity of a protostar is from accretion,
\begin{eqnarray}
   L_\star&\simeq&\frac{fGM_\star\dot{M}_\star}{R_\star}\,,
   \label{star_lum}
\end{eqnarray}
where, $f=0.75$ is the fraction of the accreted material's gravitational energy that is radiated from the surface of the protostar \citep[the rest is presumed to be removed by radiation, and by bipolar jets and outflows; ][]{OKMK09}; $M_\star$ is the mass of the protostar; $\dot{M}_\star$ is the rate of accretion onto the protostar; and $R_\star=3\,\mathrm{R}_{\odot}$ is the approximate radius of a protostar \citep{PS93}.

In the sequel, each set of initial conditions is used to run three simulations. In the first, radiative feedback from sinks is neglected (\verb|NRF|). In the second, material assimilated by a sink is presumed to be accreted instantaneously by the central protostar, and therefore delivers continuous radiative feedback (\verb|CRF|).  In the third, material assimilated by a sink is assumed to accumulate in an inner accretion disc, and then accrete onto the central star in a burst, giving episodic radiative feedback (\verb|ERF|). The \verb|NRF| simulations are performed purely for comparison, and our main concern is with the difference between continuous and episodic radiative feedback.

In order for the disc around a protostar to fragment gravitationally, at radius $R$, two criteria must be fulfilled. First, gravity must be able to overcome thermal and centrifugal support,
\begin{eqnarray}
\Sigma(R)&\ga&\frac{c(R)\kappa(R)}{\pi G}\,,
\end{eqnarray}
where $\Sigma(R)$ is the local surface-density, $c(R)$ the sound speed, and $\kappa(R)$ the epicyclic frequency; \citep{T64}. Second, the condensing fragment must cool fast enough to avoid undergoing an adiabatic bounce and then being sheared apart, i.e.
\begin{eqnarray}
   t_\textsc{cool}&\la&t_\textsc{orb}\,,
\end{eqnarray}
where $t_\textsc{cool}$ is the gas cooling time, and $t_\textsc{orb}$ is the orbital period \citep{Gam01}. Simulations that include instantaneous accretion and continuous radiative feedback \citep[e.g.][]{B09b,K06,KCKM10,OKMK09,O10,UME10} find that protostellar discs are  so warm that they cannot fragment and therefore do not spawn low-mass H-burning stars or brown dwarfs.

However, there is observational evidence suggesting that accretion onto a protostar may sometimes be episodic, i.e. it may occur intermittently, in short, intense bursts \citep[e.g.][]{H77,D78,R89,HK96,GAR08}. In particular, FU Ori-type stars \citep[e.g.][]{H77,GAR08,PSMB10,G11,PKG13} exhibit large increases in their brightness that last for between a few and a few hundred years. The physical cause of episodic accretion is uncertain. The critical issue is the duty-cycle, specifically the duration of the low-luminosity phase: 
if this is longer than the cooling timescale, typically a few times $10^3\,\mathrm{yrs}$, the disc can cool and fragment.

For the purpose of the simulations presented here we adopt the phenomenological \verb|ERF| model of \citet{SWH11,SWH12} (hereafter SWH11), which is based on the calculations by \citet{ZHG09,ZHG10} and \citet{Z10}. In the outer disc (outiside the sink), angular momentum is redistributed by gravitational torques, allowing material to spiral inwards and into the sink. However, inside the sink the material near the protostar is so hot that the inner disc becomes gravitationally stable, and there are no gravitational torques to transport angular momentum outwards. Consequently this material cannot spiral inwards any further, and instead accumulates in the inner disc. This continues until the accumulated material becomes so hot that it is ionised thermally, sufficiently to couple to the magnetic field. The magneto-rotational instability (MRI) then cuts in and the resulting torques enable material to spiral inwards and onto the star, giving a short burst of very high luminosity. The downtime -- during which material is accumulating in the inner disc, the luminosity is low, and the outer disc may be able to fragment -- is given by
\begin{equation}
t_\textsc{down}\sim 13\,{\rm kyr}\left(\frac{M_\star}{0.2\,\mathrm{M}_{\odot}}\right)^{2/3}\left(\frac{\dot{M}_\textsc{sink}}{10^{-5}\,\mathrm{M}_{\odot}\mathrm{yr}^{-1}}\right)^{-8/9}\,;
\end{equation}
$\dot{M}_\textsc{sink}$ is the rate at which material flows into the sink.


\section{Results}\label{results}%

Each core is evolved for $2\times10^{5}\,\mathrm{yrs}$\,, which is longer than the freefall time of the least-dense core. Of the 100 cores sampled from the lognormal distribution, 60 collapsed to form at least one star; note that the simulations with different feedback are identical up until this point. The star formation efficiencies and numbers of protostars formed per core are given in Table \ref{onlinematerial} in the online material. In the figures we use the naming convention \texttt{NUM\_FBK}, where \texttt{NUM} is the core number (identifying the initial conditions) and \texttt{FBK} is the feedback type.

\begin{figure}
   \centering

   \includegraphics[width=\columnwidth]{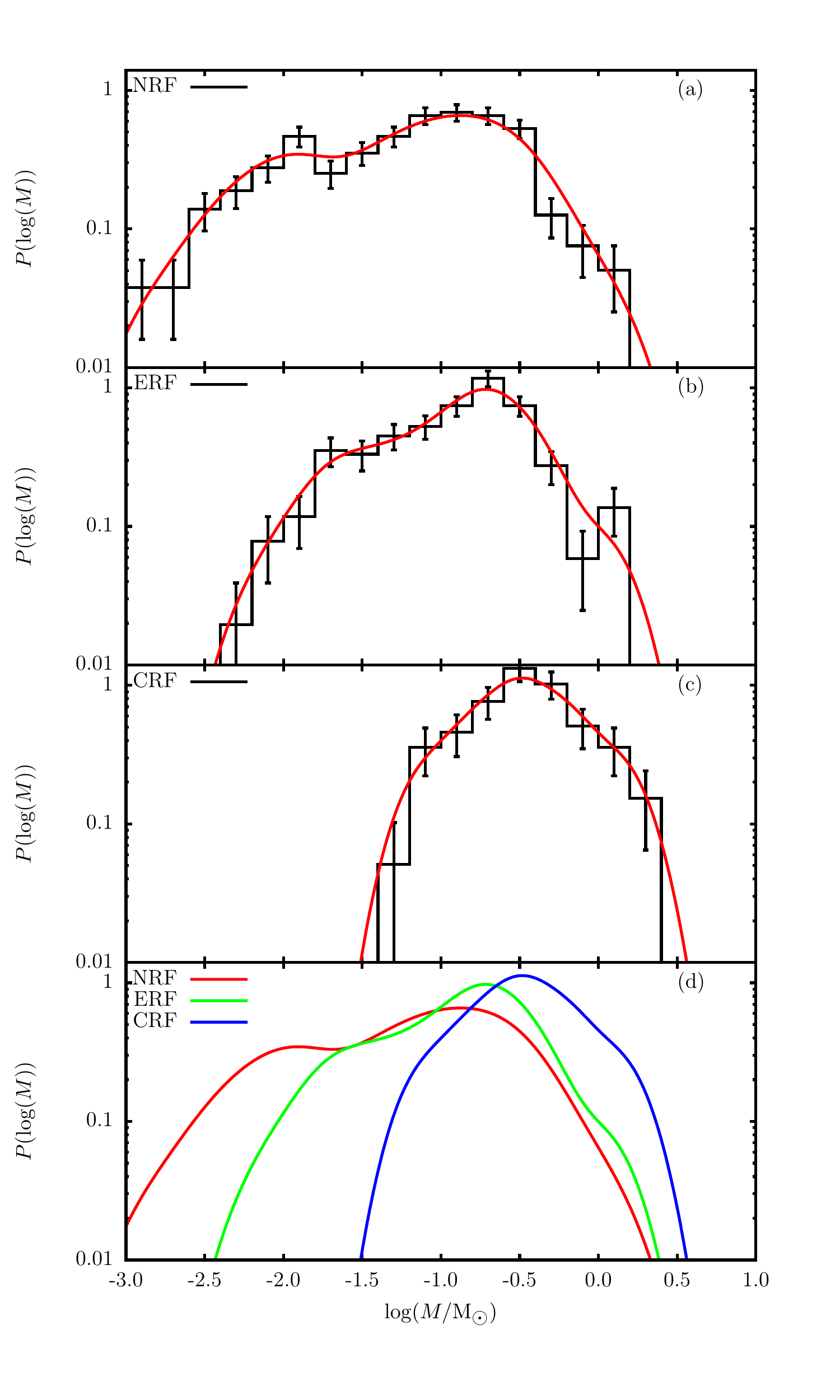}

   \caption{Protostar mass functions. The top three frames (a,b,c) give the mass functions from simulations with, respectively, \texttt{NRF}, \texttt{ERF} and \texttt{CRF}. The black histograms have bins equally spaced in $\log(M)$ and the red lines are kernel smoothed density functions. The bottom frame (d) shows the smoothed \texttt{NRF}, \texttt{ERF} and \texttt{CRF} density functions on a single plot.}
   \label{masses1}
\end{figure}

\subsection{Protostellar masses}%

Fig. \ref{masses1} shows, as histograms and smoothed probability densities\footnotemark, the protostellar mass functions from the simulations. Both the \verb|NRF| and \verb|ERF| simulations deliver a wide range of masses from a few Jupiter masses to between one and two solar masses. The \verb|CRF| simulations deliver roughly the same upper mass limit, but no protostars below about 50 Jupiter masses. In the \verb|NRF| simulations, the mean efficiency (total mass of stars formed divided by core mass) is 71\%, with on average 3.3 stars and 3.4 brown dwarfs per core. In the \verb|ERF| simulations, the efficiency is again 71\%, with on average 3.1 stars and 1.5 brown dwarfs per core. In the \verb|CRF| simulations, the efficiency is 59\%, with on average 1.6 stars and 0.1 brown dwarfs per core.

\footnotetext{The smoothed probability density is given by
\begin{equation}
   P(\log(M))=\frac{1}{N}\sum\limits_{i=1}^N \frac{1}{\sqrt{2\uppi h^2}}\exp\left(-\frac{(\log(M)-\log(M_i))^2}{2h^2}\right)\,.
\end{equation}
In the expectation that $P(\log(M))$ is roughly lognormal, we set
\begin{equation}
   h=\left(\frac{4\hat{\sigma}^5}{3N}\right)^\frac{1}{5}\,
\end{equation}
where $\hat{\sigma}$ is the standard deviation of $\log(M)$ \citep{s98}. This is a useful method of extracting from noisy histograms large scale features that are otherwise hard to see.}

\subsubsection*{Ratio of stars to brown dwarfs}%

\citet{AMGA08} have performed a survey of star forming regions and find that low mass stars outnumber brown dwarfs by a factor
\begin{equation}
   \mathcal{A}=\frac{N(0.08\,\mathrm{M}_{\odot}<M\leq1.0\,\mathrm{M}_{\odot})}{N(0.03\,\mathrm{M}_{\odot}<M\leq0.08\,\mathrm{M}_{\odot})}=4.3\pm1.6\,.
\end{equation}
In the simulations presented here, $\mathcal{A}_\textsc{nrf}=2.2\pm 0.3$, $\mathcal{A}_\textsc{erf}=3.9\pm 0.6$ and $\mathcal{A}_\textsc{crf}=17\pm 8$. This confirms that \verb|NRF| produces too many brown dwarfs, and suggests that \verb|CRF| produces too few. \verb|ERF| appears to produce about the right number.

\subsubsection*{Comparison to the IMF}%

\begin{figure}
   \centering
   \includegraphics[width=\columnwidth]{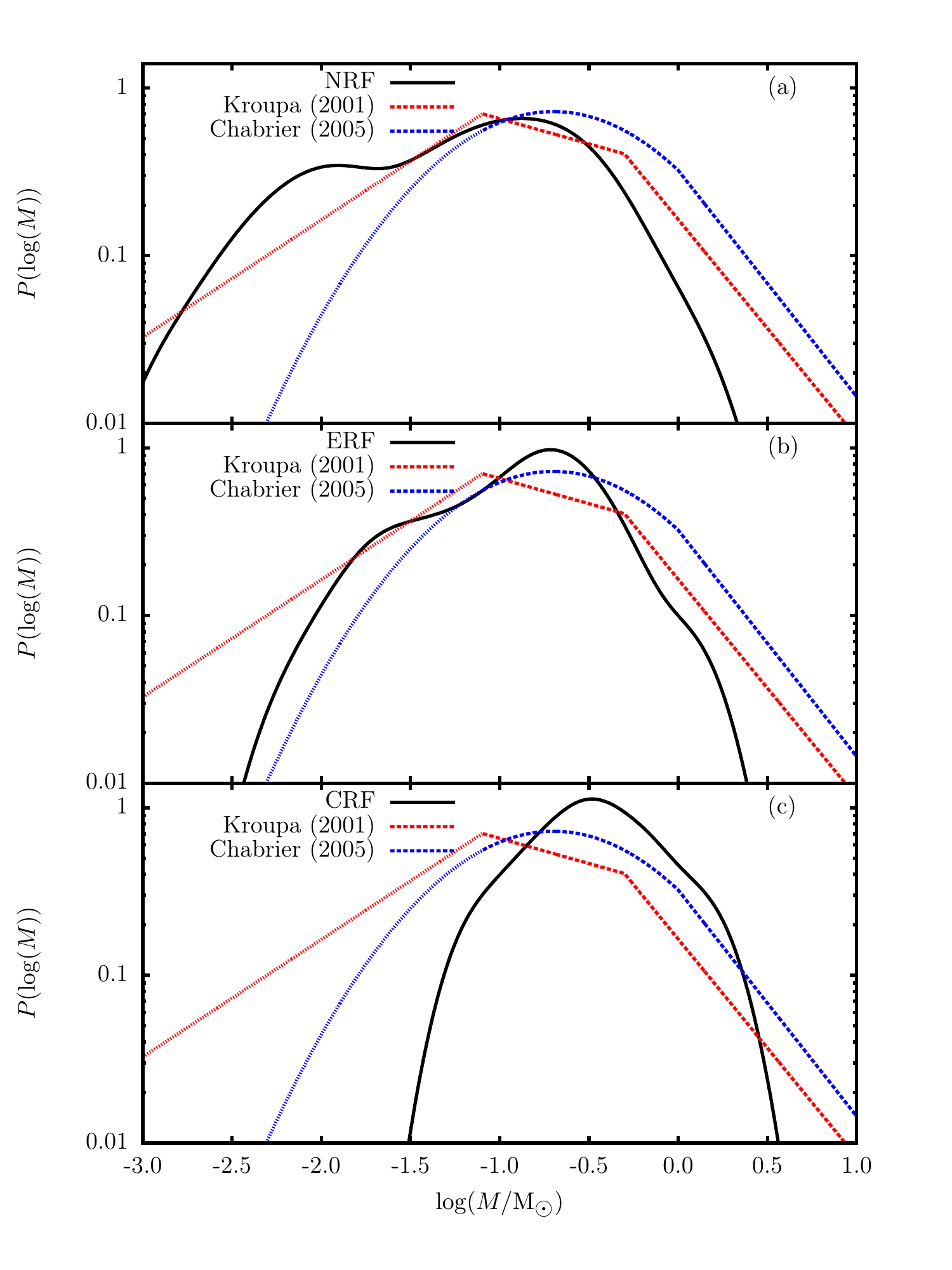}
   \caption{As Fig. \ref{masses1}, but here the simulated protostellar mass functions (in black) are compared with the K01 and C05 IMFs (in red and blue respectively). The three frames show protostellar mass function for simulations with (a) \texttt{NRF}, (b) \texttt{ERF}, and (c) \texttt{CRF}. The dashed parts of the K01 and C05 IMFs are in the stellar mass regime ($M\geq0.08\,\mathrm{M_{\odot}}$) and the dotted parts are in the brown dwarf mass regime ($M<0.08\,\mathrm{M_{\odot}}$), which is not as well constrained as the stellar regime. All mass functions are normalised so that $\int_{-\infty}^{+\infty}P(\log(M))\,\mathrm{d}\log(M)=1$.}
   \label{masses2}
\end{figure}

In Fig. \ref{masses2}, we have plotted the mass distributions from Fig. \ref{masses1} with the Kroupa (2001; K01) and Chabrier (2005; C05) fits to the IMF. The K01 IMF peaks at $M_\textsc{peak}^\textsc{K01}=0.08\,\mathrm{M}_{\odot}$, and the C05 IMF at $M_\textsc{peak}^\textsc{C05}=0.2\,\mathrm{M}_{\odot}$. In the simulations with \verb|NRF|, \verb|ERF| and \verb|CRF|, the mass distributions peak at $M_\textsc{peak}^\textsc{nrf}\approx0.1\,\mathrm{M}_{\odot}$, $M_\textsc{peak}^\textsc{erf}\approx0.2\,\mathrm{M}_{\odot}$ and $M_\textsc{peak}^\textsc{crf}\approx0.4\,\mathrm{M}_{\odot}$, respectively. The peaks of the \verb|NRF| and \verb|ERF| mass distributions are in good agreement with those of K01 and C05 respectively; the peak of the \verb|CRF| mass distribution appears to be rather too high. There are no high-mass cores in Ophiuchus, and therefore none of the simulated IMFs match the observed IMFs at high masses.

\subsubsection*{Summary: masses}%

For masses less than $\sim\!1\,\mathrm{M}_\odot$, the \verb|ERF| simulations deliver a protostellar mass function which is compatible with observed IMFs. In contrast, the \verb|CRF| simulations appear to produce too few low-mass H-burning stars and brown dwarfs. In the \verb|ERF| simulations each core produces on average $N_{_{\rm ERF}}\approx4.6$ stars with a star formation efficiency of $\eta_{_{\rm ERF}}\approx0.7$. This is in agreement with the statistical inference of \citet{HWGW13} that values of $N=4.3\pm0.4$ and $\eta=1.0\pm0.3$ are required to reproduce the observed multiplicity statistics ($\eta>1$ simply implies that the cores accrete additional mass as they form stars). In the \verb|CRF| simulations the number of stars formed is significantly lower, $N_{_{\rm CRF}}=1.7$, and the efficiency is a little lower, $\eta_{_{\rm CRF}}=0.59$.

\subsection{Fragmentation}%

\subsubsection*{Mechanisms}%

\begin{figure}
   \centering
   \includegraphics[angle=-90,width=\columnwidth]{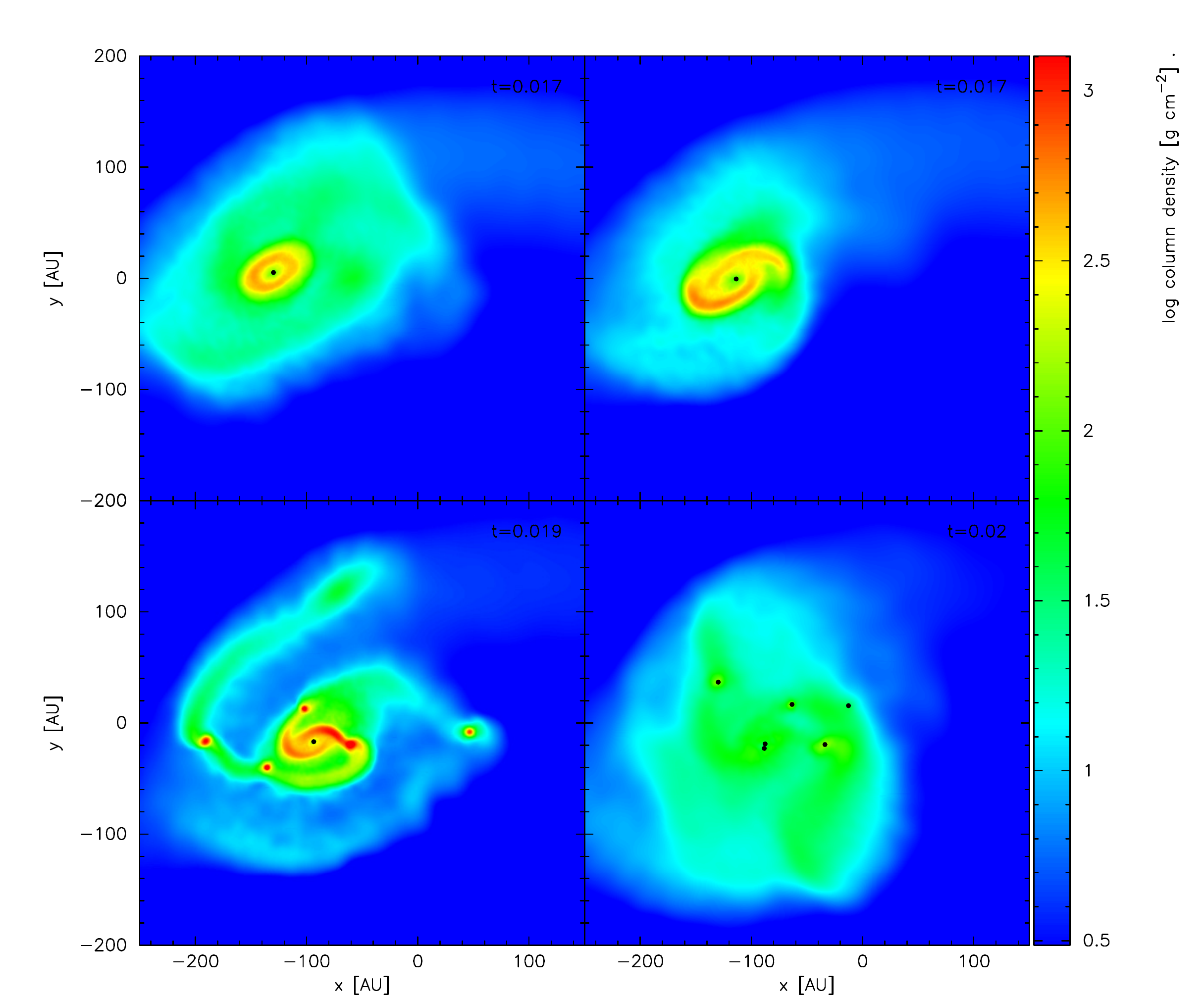}
   \caption[Simulation snapshots showing disc fragmentation.]{Column density plots from simulation \texttt{002\_ERF} showing disc fragmentation. From left to right, then top to bottom, the times are $t=17,\,18,\,19\,{\rm and}\,20\,{\rm kyr}$. This core fragments into three stars and three brown dwarfs. This figure and subsequent simulation snapshots are rendered using \textsc{Splash} \citep{P07}.}
   \label{disc}
\end{figure}

In many simulations, the core collapses to form a central protostar surrounded by an accretion disc. If the disc is sufficiently massive and cold, it fragments to form additional protostars, typically with masses smaller than the central protostar. An example of this is shown in Fig. \ref{disc}. We see that the time scale of disc fragmentation is of order $10^3\,\mathrm{yrs}$. This mechanism is well documented \citep[e.g.][]{SHW07,SW08,SW09a,SW09b,TKGSW,SMWA11} and has been shown to reproduce the observed properties of stars and brown dwarfs rather well \citep[e.g.][]{SW09a}.

\begin{figure}
   \centering
   \includegraphics[angle=-90,width=\columnwidth]{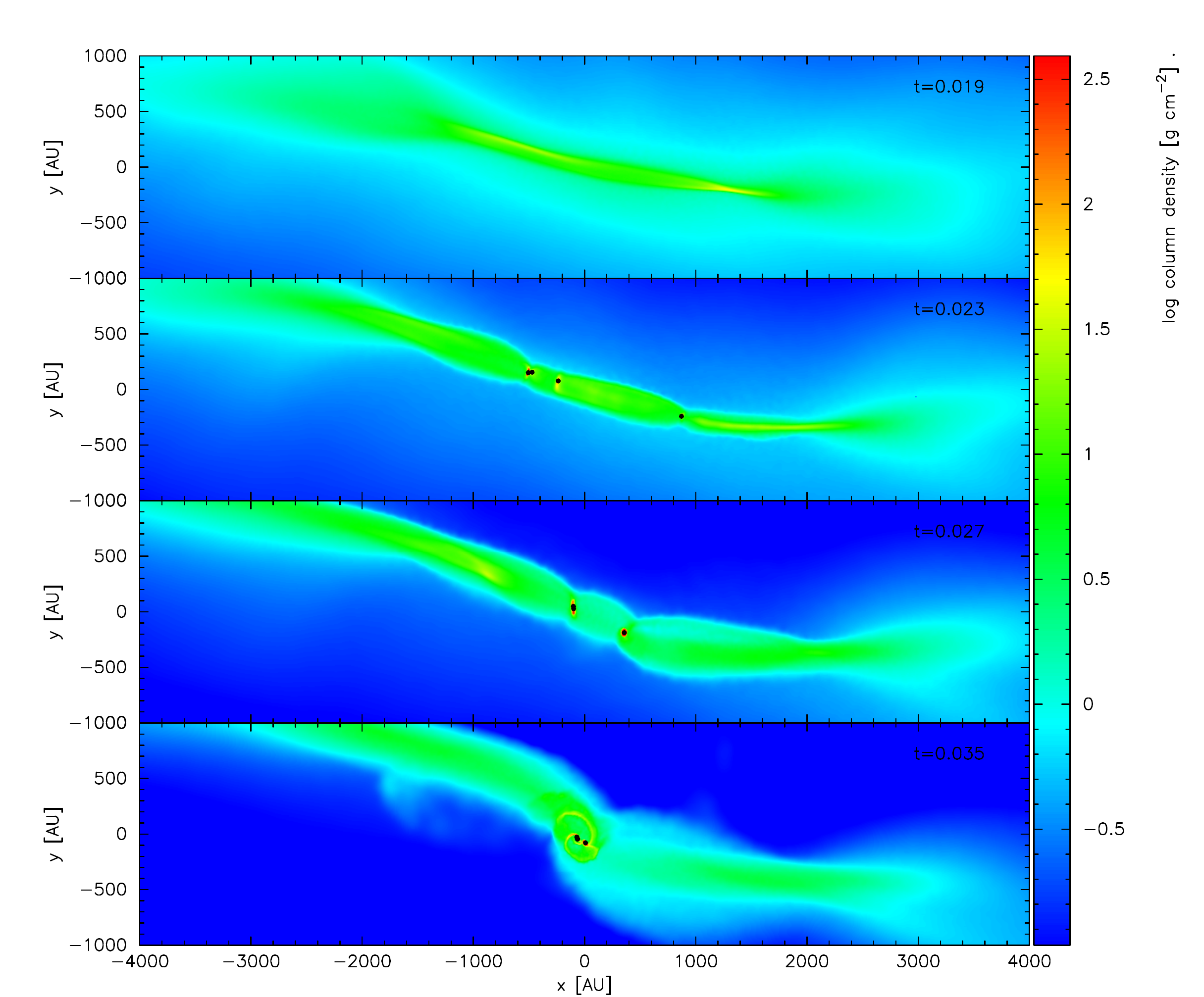}
   \caption[Simulation snapshots showing filament fragmentation.]{Column density plots from simulation \texttt{098\_ERF} showing the filament fragmentation. From top to bottom, the times are $t=19,\,23,\,27\,{\rm and}\,36\,{\rm kyrs}$. This core fragments into eight stars and two brown dwarfs.}
   \label{filament}
\end{figure}

In most simulations the central regions are fed by filamentary streams of material, and in some cases these filaments fragment before a central protostar forms; this also occurs on time scales of order $10^3\,\mathrm{yrs}$. An example of this is shown in Fig. \ref{filament}.  This core has a relatively high mass ($M\!=\!3.3\,\mathrm{M}_{\odot}$) and a low ratio of turbulent to gravitational energy ($\alpha_{_\textsc{turb}}=0.07$). The core collapses into a filament because it is particularly elongated, with its longest axis more than twice the length of the other two. The filament then spawns ten protostars. This confirms the results of \citet{GWW04}, who showed that low levels of turbulence are sufficient to fragment spherical cores.

\subsubsection*{Feedback effects}%

\begin{figure}
   \centering
   \includegraphics[angle=-90,width=\columnwidth]{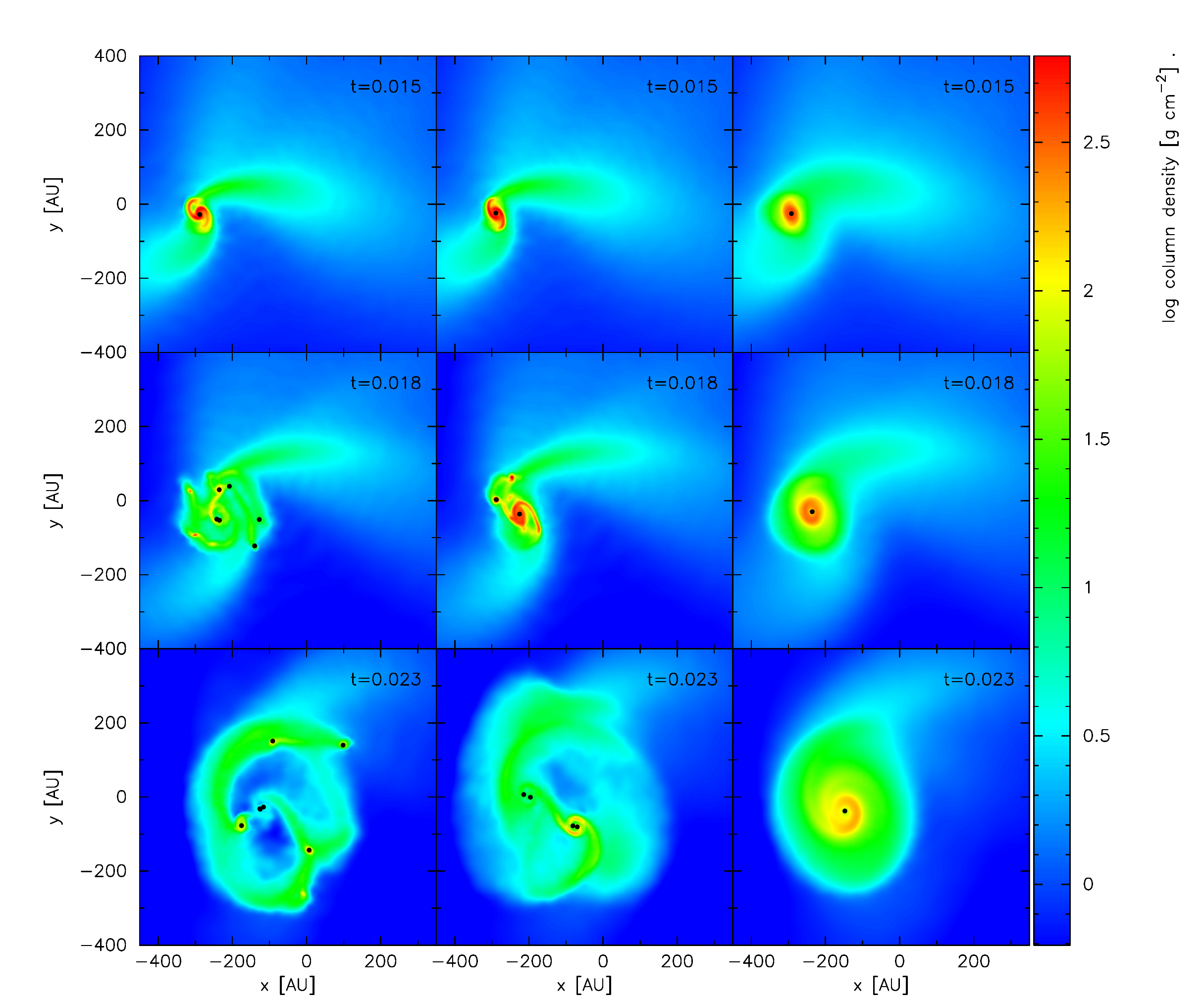}
   \caption[Simulation snapshots comparing the effects of different accretion feedback models.]{Column density plots from the \texttt{022\_XXX} simulations. The left-hand column gives snapshots from \texttt{022\_NRF}, the middle column from \texttt{022\_ERF}, and the right-hand column from \texttt{022\_CRF}. From top to bottom, the times are $t=15,\,18\,{\rm and}\,23\,\mathrm{kyr}$. \texttt{022\_NRF} forms four stars and nine brown dwarfs, \texttt{022\_ERF} forms four stars, and \texttt{022\_CRF} forms a single star.}
   \label{feedback}
\end{figure}

Fig. \ref{feedback} shows a comparison of column density plots from simulations using different feedback mechanisms. There are two main differences between the protostellar mass distributions from the \verb|NRF| and \verb|ERF| simulations. First, the main peak shifts from $\sim\!0.1\mathrm{M_\mathrm{\odot}}$ in the \verb|NRF| simulations, to $\sim\!0.2\mathrm{M_\mathrm{\odot}}$ in the \verb|ERF| simulations. Second, in the \verb|NRF| simulations there is a second peak around $\sim 0.01\mathrm{M_\mathrm{\odot}}$. The main peaks are populated by protostars that form from core collapse. The secondary peak in the \verb|NRF| simulations arises from prolific disc fragmentation. This produces an abundance of low mass objects, many of which are ejected from the core \citep[e.g.][]{BBB02}. Once ejected, accretion ceases and their masses are frozen at low values.

In the \verb|ERF| simulations, high-luminosity outbursts periodically raise the temperature of the gas surrounding a protostar, stabilising it against gravitational fragmentation. Initially, these outbursts occur at intervals of $\Delta t_\textsc{down}\sim10^4\,\mathrm{yrs}$, whereas gravitational instabilities develop on times scales of $\Delta t_\textsc{gi}\sim10^3\,\mathrm{yrs}$; therefore there are windows of opportunity for fragmentation between outbursts. However, as further protostars form, the intervals between outbursts decrease until eventually $\Delta t_\textsc{down}\la \Delta t_\textsc{gi}$. Thereafter, further star formation is inhibited by regular luminosity outbursts. Thus \verb|ERF|  allows fragmentation to occur, but then regulates the process after a few protostars have formed. Because fewer low-mass protostars are formed, there are fewer dynamical ejections, and more mass is available to accrete onto the central protostar \citep[cf.][]{GFB12}, so the main peak is enhanced and the secondary peak is removed.

In the \verb|CRF| simulations, once the first protostar has formed, the remaining gas is continuously heated and very little further fragmentation occurs, so these simulations usually form single protostars or binaries. Fig. \ref{feedback} shows an example of a core evolving with \verb|NRF|, \verb|ERF| and \verb|CRF|; the usual trend is that the most protostars are formed with \verb|NRF|, followed by \verb|ERF|, then \verb|CRF|.

\subsubsection*{Number of objects formed}\label{nobjects}%

\begin{figure}
   \centering
   \includegraphics[width=\columnwidth]{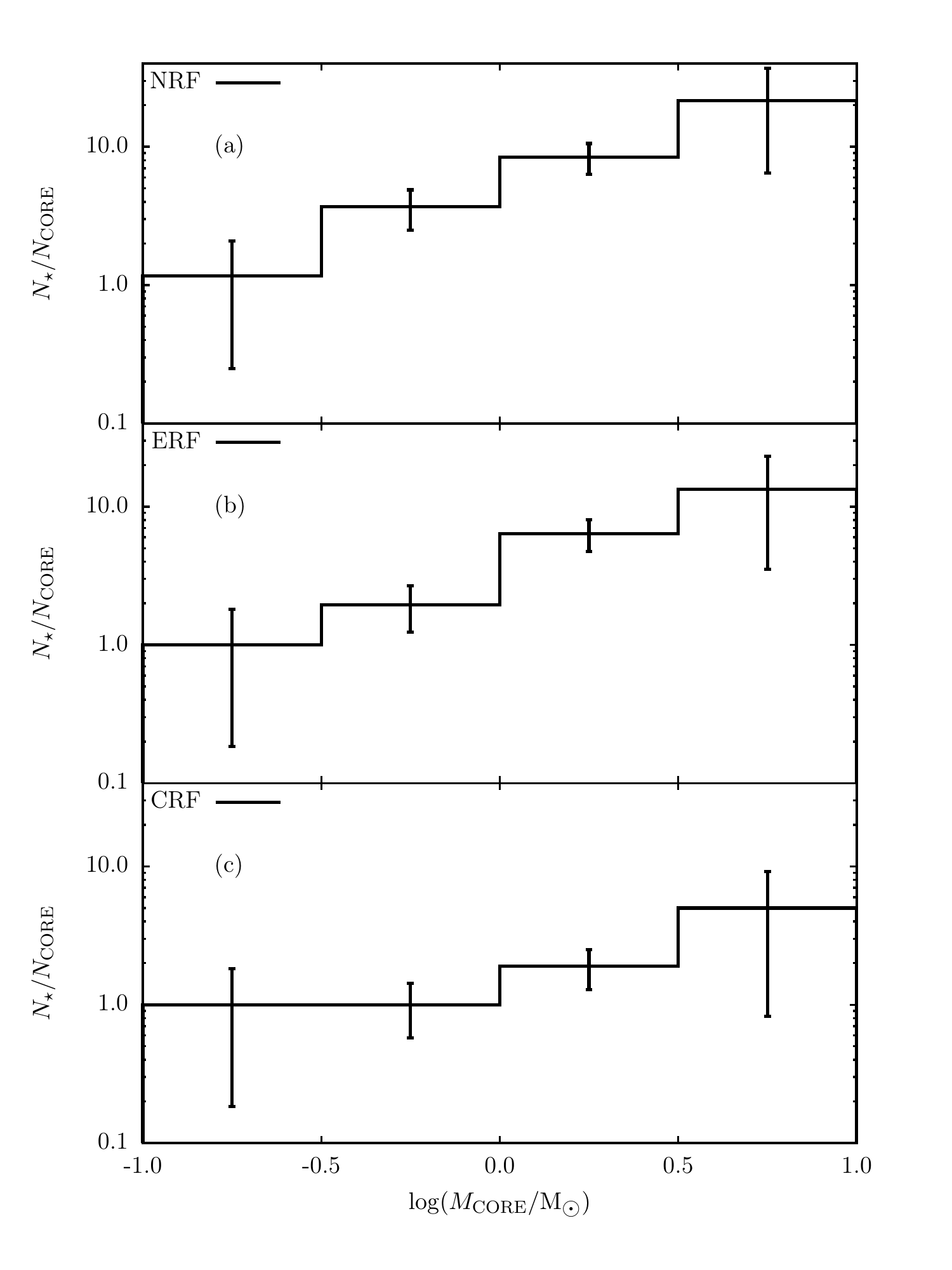}
   \caption[The number of protostars formed per core as a function of core mass.]{The number of protostars formed per core as a function of core mass. Error bars reflect the Poisson uncertainties in the numbers of cores and sinks, per bin. From top to bottom, simulations invoking (a) \texttt{NRF}, (b) \texttt{ERF} and (c) \texttt{CRF}.}
   \label{N0}
\end{figure}

Fig. \ref{N0} shows that the number of protostars formed per core is correlated with the core mass. Cores with masses between $0.1\,\mathrm{M_{\odot}}$ and $0.3\,\mathrm{M_{\odot}}$ typically form single protostars, whereas cores with masses between $3\,\mathrm{M_{\odot}}$ and $10\,\mathrm{M_{\odot}}$ typically produce more than ten protostars with \verb|ERF|, and about five with \verb|CRF|. For convenience, we will refer to cores with $M_\textsc{core}\leq1\,\mathrm{M}_{\odot}$ as low-mass cores, and cores with $M_\textsc{core}>1\,\mathrm{M}_{\odot}$ as high-mass cores.

\begin{figure}
   \centering
   \includegraphics[width=\columnwidth]{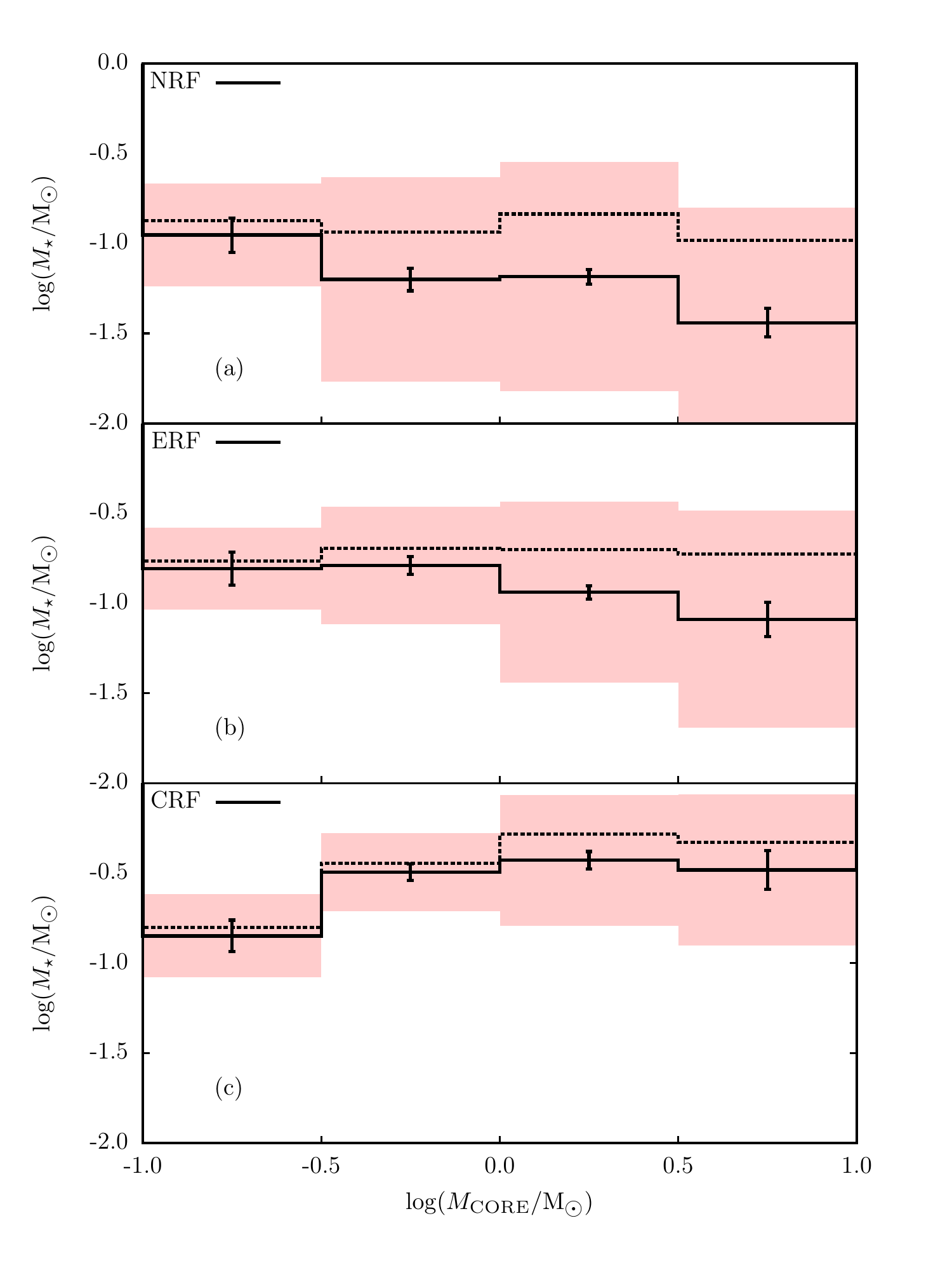}
   \caption[The distribution of protostar masses as a function of core mass.]{The distribution of protostar masses as a function of core mass. The solid line gives the geometric mean mass and standard error. The shaded area shows the geometric standard deviation. The dashed line gives the arithmetic mean protostar mass. From top to bottom, simulations invoking (a) \texttt{NRF}, (b) \texttt{ERF} and (c) \texttt{CRF}.}
   \label{mu0}
\end{figure}

Fig. \ref{mu0} shows the distribution of protostar masses as a function of core mass. In the \verb|NRF| and \verb|ERF| simulations, the arithmetic mean mass is roughly constant over the entire range of core masses, viz. $\bar{M}_\star^\textsc{nrf}=0.12\,\mathrm{M}_{\odot}$ and $\bar{M}_\star^\textsc{erf}=0.19\,\mathrm{M}_{\odot}$. Also shown on Fig. \ref{mu0} is the geometric mean of the protostar masses and its standard deviation. This shows that high-mass cores form a wider range of masses than low-mass cores. In low-mass cores, there is little spread in the protostar masses because most of the available gas goes into the central protostar. A disc may form, but it is often not massive enough to fragment. High-mass cores have sufficient gas to form more massive protostars, and there is often enough gas remaining to build a massive disc, which can then fragment into low-mass protostars. Final masses are determined by competitive accretion for the remaing core mass.

In simulations with \verb|CRF|, low-mass cores tend to form single protostars, and high-mass cores tend to form multiple protostars, but in smaller numbers than with \verb|ERF|.

\subsubsection*{Fragmentation: summary}%

As noted by \citet{SWH12}, \verb|ERF| provides a window of opportunity for disc fragmentation. As further protostars form, they too radiate and the window eventually becomes too short to allow fragmentation. This means that, unlike \verb|CRF|, \verb|ERF| permits brown dwarfs to form by disc fragmentation, but, unlike \verb|NRF|, brown dwarf formation is regulated to a level that agrees better with observation.

Irrespective of the type of feedback invoked, these simulations do not support the notion that the shape of the IMF is inherited from the shape of the CMF. The mean mass of a protostar formed in a low-mass core is approximately the same as one formed in a high-mass core. The difference is that high-mass cores tend to produce more protostars, roughly in proportion to their mass.

Competitive accretion occurs in our simulations, but only on the scale of individual cores having sufficient mass ($M_{_{\rm CORE}}\!>\!{\rm M}_{_\odot}$) to form multiple protostars.


\subsection{Limitations}\label{limitations}%

Since the most massive core in Ophiuchus has mass $\sim 5\,{\rm M}_{_\odot}$ (and the next most massive $\sim 3\,{\rm M}_{_\odot}$), and since we evolve our cores in isolation (i.e. they do not accrete extra material from their surroundings), we do not reproduce the high-mass tails of the K01 or C05 IMFs, irrespective of which feedback prescription is invoked. If core growth were included, say by virtue of the cores being embedded in filaments, we would expect the rate of growth to be quite low, $\dot{M}_{_{\rm CORE}}\la M_{_{\rm CORE}}/t_{_{\rm FF}}$. Under this circumstance the inflowing material is unlikely to spawn new protostars; instead it will sustain the reservoir of gas in the centre of the core, thereby prolonging competitive accretion. Consequently the more massive protostars will be the major beneficiaries, and the IMF will broaden towards higher masses.

\section{Summary and conclusions}\label{conclusions}%

We have performed a large ensemble of simulations of prestellar cores collapsing and fragmenting to form protostars. The initial conditions are informed by observations of cores in Ophiuchus, and reproduce the observed distributions of mass, size, aspect ratio and velocity dispersion. The mass resolution is high enough to resolve the opacity limit, $M_\textsc{o}\sim3\times10^{-3}\,\mathrm{M}_{\odot}$, with $\sim\!300$ SPH particles.

Simulations that invoke continuous radiative feedback produce an IMF that peaks at too high a mass ($\sim 0.4\,{\rm M}_{_\odot}$) and contains very few brown dwarfs. Simulations that invoke episodic radiative feedback produce an IMF that peaks at the right mass ($\sim 0.2\,{\rm M}_{_\odot}$) and contains roughly the correct proportion of brown dwarfs.

The masses of the protostars are not strongly correlated with the masses of the cores in which they form -- and therefore the shape of the IMF is not inherited from the shape of the CMF. Rather, low-mass cores ($M_{_{\rm CORE}}<{\rm M}_{_\odot}$) tend to spawn just one or two protostars. In contrast, high-mass cores tend to spawn small-$N$ clusters of stars, with a mean mass that is comparable with the mean mass for low-mass cores, but a greater spread of masses. This is because, in a high-mass core, (a) there is a greater chance that the first protostar to form then acquires a massive disc that fragments to form low-mass companions, and (b) there is a larger flow of material into the center, and so the protostars there can grow via competitive accretion.

In a companion paper (Lomax et al. in preparation) we will present and evaluate the multiplicity statistics of the protostars formed in these simulations. 

\bibliographystyle{mn2e}\bibliography{refs}%

\section*{Acknowledgements}

We gratefully acknowledge the support of the STFC, via a doctoral training account (OL) and a rolling grant (DS, OL \& APW; ST/K00926/1). SW acknowledges support from the DFG priority programme "Physics of the ISM".

%
%
%
%
%

\newpage
\onecolumn

\begin{landscape}
\section*{Online Material}
\LTcapwidth=\textwidth
\begin{longtable}{ccccccccc | ccc | ccc | ccc}

\caption{Initial conditions and results of simulations. Column 1 gives the core number; column 2 gives the total mass of the core, column 3 gives the one-dimensional non-thermal velocity dispersion assigned to the core; column 4 gives the effective radius of the core where $R_\textsc{core}=(A_\textsc{core}B_\textsc{core}C_\textsc{core})^{\frac{1}{3}}$ and $A_\textsc{core}$, $B_\textsc{core}$ and $C_\textsc{core}$ are the semi-axes of the core; column 5 gives the longest axis of the core; column 6 gives the intrinsic aspect ratio between the longest and intermediate axis of the core; column 7 gives the intrinsic aspect ratio between the longest and shortest axis of the core; column 8 gives the average density of the core; column 9 gives the notional free-fall time of the core where $t_\textsc{ff}=(\frac{3\uppi}{32\,G\,\bar{\rho}})^{\frac{1}{2}}$; columns 10, 11 and 12 give the total sink mass, the number of stellar-mass protostars (i.e. $M>0.08\,\mathrm{M}_\odot$) and the number of brown dwarf-mass protostars (i.e. $M\leq0.08\,\mathrm{M}_\odot$) formed with no radiative feedback from protostars; columns 13, 14 and 15 give the same quantities when there is episodic radiative feedback; columns 16, 17 and 18 give the quantities when there is continuous radiative feedback.} \\
\hline
&&&&&&&&&&&&&&&&&\\
&&&&&&&&& \multicolumn{3}{c}{NRF} \vline & \multicolumn{3}{c}{ERF} \vline & \multicolumn{3}{c}{CRF} \\
Core N & $M_\textsc{core}$ & $\sigma_\textsc{nt}$ & $R_\textsc{core}$ & $A_\textsc{core}$ & $B/A$ & $C/A$ & $\bar{\rho}$ & $t_\textsc{ff}$ & $\sum{M_\star}$ & $N_{\star}$ & $N_\textsc{bd}$ & $\sum{M_\star}$ & $N_{\star}$ & $N_\textsc{bd}$ & $\sum{M_\star}$ & $N_{\star}$ & $N_\textsc{bd}$ \\
& $(\mathrm{M}_\odot)$ & $(\mathrm{km\,s^{-1}})$ & $(\mathrm{AU})$ & $(\mathrm{AU})$ & & & $(10^{-17}\,\mathrm{g\,cm^{-3}})$ & $(\mathrm{Myr})$ & $(\mathrm{M}_\odot)$ & & & $(\mathrm{M}_\odot)$ & & & $(\mathrm{M}_\odot)$ & & \\
&&&&&&&&&&&&&&&&&\\
\hline
\endfirsthead\\

\caption{Initial conditions and results of simulations.}\\
\hline
&&&&&&&&&&&&&&&&&\\
&&&&&&&&& \multicolumn{3}{c}{NRF} \vline & \multicolumn{3}{c}{ERF} \vline & \multicolumn{3}{c}{CRF} \\
Core N & $M_\textsc{core}$ & $\sigma_\textsc{nt}$ & $R_\textsc{core}$ & $A_\textsc{core}$ & $B/A$ & $C/A$ & $\bar{\rho}$ & $t_\textsc{ff}$ & $\sum{M_\star}$ & $N_{\star}$ & $N_\textsc{bd}$ & $\sum{M_\star}$ & $N_{\star}$ & $N_\textsc{bd}$ & $\sum{M_\star}$ & $N_{\star}$ & $N_\textsc{bd}$ \\
& $(\mathrm{M}_\odot)$ & $(\mathrm{km\,s^{-1}})$ & $(\mathrm{AU})$ & $(\mathrm{AU})$ & & & $(10^{-17}\,\mathrm{g\,cm^{-3}})$ & $(\mathrm{Myr})$ & $(\mathrm{M}_\odot)$ & & & $(\mathrm{M}_\odot)$ & & & $(\mathrm{M}_\odot)$ & & \\
&&&&&&&&&&&&&&&&&\\
\hline
\endhead\\

\hline
\endfoot

1 & 0.478 & 0.066 & 2360 & 5045 & 0.333 & 0.308 & 0.516 & 0.029 & 0.331 & 1 & 2 & 0.283 & 1 & 0 & 0.285 & 1 & 0 \\
2 & 0.809 & 0.135 & 2222 & 2722 & 0.906 & 0.601 & 1.045 & 0.021 & 0.766 & 3 & 10 & 0.597 & 3 & 3 & 0.563 & 1 & 0 \\
3 & 1.881 & 0.122 & 6945 & 7728 & 0.947 & 0.766 & 0.080 & 0.075 & 1.709 & 4 & 3 & 1.393 & 5 & 3 & 1.223 & 1 & 0 \\
4 & 0.096 & 0.172 & 2718 & 3887 & 0.657 & 0.520 & 0.068 & 0.081 & 0.000 & 0 & 0 & 0.000 & 0 & 0 & 0.000 & 0 & 0 \\
5 & 0.169 & 0.090 & 2474 & 3961 & 0.538 & 0.453 & 0.159 & 0.053 & 0.000 & 0 & 0 & 0.000 & 0 & 0 & 0.000 & 0 & 0 \\
6 & 0.110 & 0.069 & 2666 & 4320 & 0.508 & 0.463 & 0.082 & 0.073 & 0.000 & 0 & 0 & 0.000 & 0 & 0 & 0.000 & 0 & 0 \\
7 & 1.425 & 0.179 & 8054 & 13833 & 0.525 & 0.376 & 0.039 & 0.107 & 0.000 & 0 & 0 & 0.000 & 0 & 0 & 0.000 & 0 & 0 \\
8 & 2.999 & 0.098 & 13347 & 24006 & 0.586 & 0.294 & 0.018 & 0.157 & 1.190 & 6 & 4 & 0.923 & 5 & 4 & 1.015 & 6 & 0 \\
9 & 0.540 & 0.132 & 3281 & 4193 & 0.864 & 0.554 & 0.217 & 0.045 & 0.171 & 1 & 0 & 0.174 & 1 & 0 & 0.171 & 1 & 0 \\
10 & 2.358 & 0.176 & 3556 & 6121 & 0.508 & 0.386 & 0.744 & 0.024 & 2.142 & 5 & 8 & 2.145 & 2 & 1 & 1.437 & 1 & 0 \\
11 & 1.307 & 0.160 & 3677 & 4087 & 0.917 & 0.794 & 0.373 & 0.034 & 1.198 & 7 & 3 & 1.207 & 8 & 1 & 1.156 & 4 & 0 \\
12 & 0.164 & 0.110 & 1446 & 2581 & 0.510 & 0.345 & 0.769 & 0.024 & 0.000 & 0 & 0 & 0.000 & 0 & 0 & 0.000 & 0 & 0 \\
13 & 0.088 & 0.023 & 1365 & 1611 & 0.975 & 0.625 & 0.492 & 0.030 & 0.000 & 0 & 0 & 0.000 & 0 & 0 & 0.000 & 0 & 0 \\
14 & 1.532 & 0.268 & 4034 & 5119 & 0.948 & 0.516 & 0.331 & 0.037 & 0.953 & 4 & 4 & 0.871 & 4 & 2 & 0.889 & 3 & 1 \\
15 & 1.160 & 0.203 & 8211 & 20024 & 0.306 & 0.225 & 0.030 & 0.122 & 0.000 & 0 & 0 & 0.000 & 0 & 0 & 0.000 & 0 & 0 \\
16 & 1.606 & 0.140 & 6555 & 9715 & 0.629 & 0.488 & 0.081 & 0.074 & 1.074 & 6 & 3 & 1.103 & 5 & 2 & 0.589 & 1 & 0 \\
17 & 0.330 & 0.063 & 3418 & 5890 & 0.566 & 0.345 & 0.117 & 0.062 & 0.000 & 0 & 0 & 0.000 & 0 & 0 & 0.000 & 0 & 0 \\
18 & 0.368 & 0.127 & 2653 & 3084 & 0.811 & 0.785 & 0.279 & 0.040 & 0.201 & 1 & 0 & 0.197 & 1 & 0 & 0.194 & 1 & 0 \\
19 & 0.230 & 0.092 & 859 & 915 & 0.930 & 0.892 & 5.137 & 0.009 & 0.212 & 1 & 0 & 0.211 & 1 & 0 & 0.203 & 1 & 0 \\
20 & 0.889 & 0.085 & 2637 & 4919 & 0.704 & 0.219 & 0.687 & 0.025 & 0.789 & 4 & 1 & 0.779 & 3 & 0 & 0.629 & 1 & 0 \\
21 & 3.020 & 0.271 & 2952 & 3839 & 0.943 & 0.482 & 1.665 & 0.016 & 2.836 & 5 & 4 & 2.909 & 9 & 10 & 1.947 & 1 & 0 \\
22 & 0.951 & 0.125 & 2342 & 3427 & 0.737 & 0.433 & 1.049 & 0.021 & 0.858 & 4 & 9 & 0.849 & 4 & 0 & 0.535 & 1 & 0 \\
23 & 1.427 & 0.072 & 2304 & 2551 & 0.865 & 0.852 & 1.655 & 0.016 & 1.404 & 1 & 3 & 1.321 & 1 & 0 & 1.331 & 1 & 0 \\
24 & 0.231 & 0.185 & 1495 & 2676 & 0.729 & 0.239 & 0.979 & 0.021 & 0.000 & 0 & 0 & 0.000 & 0 & 0 & 0.000 & 0 & 0 \\
25 & 0.104 & 0.118 & 557 & 1012 & 0.465 & 0.359 & 8.567 & 0.007 & 0.068 & 0 & 1 & 0.067 & 0 & 1 & 0.059 & 0 & 1 \\
26 & 2.016 & 0.119 & 5326 & 9700 & 0.411 & 0.403 & 0.189 & 0.048 & 1.322 & 5 & 5 & 1.124 & 4 & 2 & 0.649 & 1 & 0 \\
27 & 3.439 & 0.166 & 4715 & 8225 & 0.452 & 0.417 & 0.465 & 0.031 & 3.053 & 7 & 3 & 2.858 & 5 & 7 & 2.597 & 3 & 0 \\
28 & 1.137 & 0.061 & 9459 & 10348 & 0.899 & 0.850 & 0.019 & 0.152 & 0.000 & 0 & 0 & 0.000 & 0 & 0 & 0.000 & 0 & 0 \\
29 & 1.763 & 0.187 & 1927 & 4687 & 0.506 & 0.137 & 3.496 & 0.011 & 1.622 & 3 & 13 & 1.679 & 5 & 0 & 1.483 & 3 & 0 \\
30 & 0.176 & 0.090 & 1007 & 1089 & 0.913 & 0.867 & 2.438 & 0.013 & 0.000 & 0 & 0 & 0.000 & 0 & 0 & 0.000 & 0 & 0 \\
31 & 0.767 & 0.087 & 3722 & 8278 & 0.307 & 0.296 & 0.211 & 0.046 & 0.531 & 3 & 0 & 0.508 & 3 & 2 & 0.389 & 1 & 0 \\
32 & 0.202 & 0.044 & 4095 & 6312 & 0.961 & 0.284 & 0.042 & 0.103 & 0.000 & 0 & 0 & 0.000 & 0 & 0 & 0.000 & 0 & 0 \\
33 & 0.189 & 0.084 & 2874 & 7390 & 0.292 & 0.202 & 0.113 & 0.063 & 0.000 & 0 & 0 & 0.000 & 0 & 0 & 0.000 & 0 & 0 \\
34 & 0.278 & 0.067 & 864 & 1112 & 0.780 & 0.601 & 6.111 & 0.009 & 0.256 & 1 & 0 & 0.254 & 1 & 0 & 0.249 & 1 & 0 \\
35 & 1.772 & 0.162 & 3700 & 4143 & 0.973 & 0.732 & 0.496 & 0.030 & 1.688 & 4 & 10 & 1.685 & 5 & 2 & 1.308 & 1 & 0 \\
36 & 1.698 & 0.201 & 4269 & 5263 & 0.789 & 0.676 & 0.310 & 0.038 & 1.460 & 8 & 10 & 1.450 & 7 & 0 & 1.277 & 5 & 1 \\
37 & 0.235 & 0.091 & 2885 & 3246 & 0.888 & 0.791 & 0.139 & 0.056 & 0.000 & 0 & 0 & 0.000 & 0 & 0 & 0.000 & 0 & 0 \\
38 & 1.506 & 0.141 & 2372 & 3564 & 0.563 & 0.523 & 1.600 & 0.017 & 1.492 & 3 & 5 & 1.366 & 1 & 0 & 1.379 & 1 & 0 \\
39 & 0.426 & 0.124 & 4476 & 6444 & 0.645 & 0.519 & 0.067 & 0.081 & 0.000 & 0 & 0 & 0.000 & 0 & 0 & 0.000 & 0 & 0 \\
40 & 1.240 & 0.198 & 3549 & 5082 & 0.643 & 0.530 & 0.393 & 0.034 & 1.120 & 4 & 0 & 1.097 & 4 & 1 & 0.847 & 2 & 0 \\
41 & 0.518 & 0.147 & 2829 & 3381 & 0.934 & 0.628 & 0.324 & 0.037 & 0.162 & 1 & 1 & 0.139 & 1 & 0 & 0.129 & 1 & 0 \\
42 & 0.823 & 0.083 & 5656 & 10628 & 0.485 & 0.311 & 0.065 & 0.083 & 0.000 & 0 & 0 & 0.000 & 0 & 0 & 0.000 & 0 & 0 \\
43 & 1.059 & 0.056 & 3829 & 6401 & 0.567 & 0.377 & 0.268 & 0.041 & 0.922 & 4 & 3 & 0.792 & 2 & 0 & 0.786 & 1 & 0 \\
44 & 0.287 & 0.079 & 1647 & 2194 & 0.720 & 0.588 & 0.911 & 0.022 & 0.191 & 1 & 1 & 0.161 & 1 & 0 & 0.154 & 1 & 0 \\
45 & 0.773 & 0.071 & 4893 & 7223 & 0.651 & 0.477 & 0.094 & 0.069 & 0.467 & 4 & 1 & 0.448 & 3 & 1 & 0.330 & 1 & 0 \\
46 & 0.161 & 0.118 & 1652 & 2687 & 0.487 & 0.477 & 0.508 & 0.030 & 0.000 & 0 & 0 & 0.000 & 0 & 0 & 0.000 & 0 & 0 \\
47 & 0.364 & 0.123 & 2551 & 2773 & 0.958 & 0.812 & 0.311 & 0.038 & 0.148 & 1 & 0 & 0.147 & 1 & 0 & 0.137 & 1 & 0 \\
48 & 0.138 & 0.070 & 768 & 1126 & 0.654 & 0.486 & 4.327 & 0.010 & 0.108 & 1 & 0 & 0.105 & 1 & 0 & 0.100 & 1 & 0 \\
49 & 0.281 & 0.123 & 3075 & 6261 & 0.682 & 0.174 & 0.137 & 0.057 & 0.000 & 0 & 0 & 0.000 & 0 & 0 & 0.000 & 0 & 0 \\
50 & 1.365 & 0.137 & 2792 & 2995 & 0.994 & 0.815 & 0.890 & 0.022 & 1.274 & 6 & 4 & 1.243 & 4 & 4 & 0.956 & 1 & 0 \\
51 & 2.272 & 0.115 & 6165 & 9575 & 0.745 & 0.358 & 0.138 & 0.057 & 2.083 & 4 & 2 & 2.055 & 4 & 3 & 1.606 & 1 & 0 \\
52 & 1.300 & 0.299 & 2905 & 4481 & 0.614 & 0.444 & 0.752 & 0.024 & 0.937 & 5 & 1 & 1.014 & 6 & 1 & 0.615 & 2 & 0 \\
53 & 0.399 & 0.133 & 2681 & 4405 & 0.502 & 0.449 & 0.293 & 0.039 & 0.000 & 0 & 0 & 0.000 & 0 & 0 & 0.000 & 0 & 0 \\
54 & 0.546 & 0.067 & 2600 & 3567 & 0.981 & 0.395 & 0.441 & 0.032 & 0.405 & 3 & 1 & 0.328 & 1 & 0 & 0.342 & 1 & 0 \\
55 & 0.236 & 0.094 & 1877 & 2482 & 0.864 & 0.500 & 0.505 & 0.030 & 0.000 & 0 & 0 & 0.000 & 0 & 0 & 0.000 & 0 & 0 \\
56 & 0.455 & 0.080 & 2930 & 3855 & 0.765 & 0.574 & 0.257 & 0.042 & 0.277 & 1 & 0 & 0.276 & 1 & 0 & 0.268 & 1 & 0 \\
57 & 0.212 & 0.071 & 1527 & 1867 & 0.905 & 0.605 & 0.845 & 0.023 & 0.000 & 0 & 0 & 0.000 & 0 & 0 & 0.000 & 0 & 0 \\
58 & 0.784 & 0.226 & 1763 & 2070 & 0.929 & 0.665 & 2.030 & 0.015 & 0.699 & 1 & 0 & 0.695 & 1 & 0 & 0.693 & 1 & 0 \\
59 & 0.290 & 0.035 & 4279 & 5756 & 0.778 & 0.528 & 0.052 & 0.092 & 0.000 & 0 & 0 & 0.000 & 0 & 0 & 0.000 & 0 & 0 \\
60 & 0.749 & 0.089 & 4664 & 9058 & 0.434 & 0.315 & 0.105 & 0.065 & 0.000 & 0 & 0 & 0.000 & 0 & 0 & 0.000 & 0 & 0 \\
61 & 0.229 & 0.091 & 2989 & 3776 & 0.772 & 0.642 & 0.121 & 0.060 & 0.000 & 0 & 0 & 0.000 & 0 & 0 & 0.000 & 0 & 0 \\
62 & 1.279 & 0.123 & 9222 & 10779 & 0.939 & 0.667 & 0.023 & 0.138 & 0.192 & 0 & 4 & 0.165 & 1 & 2 & 0.096 & 1 & 0 \\
63 & 1.022 & 0.158 & 2397 & 3293 & 0.928 & 0.415 & 1.053 & 0.021 & 0.927 & 6 & 3 & 0.913 & 5 & 2 & 0.584 & 1 & 0 \\
64 & 0.815 & 0.137 & 3742 & 4487 & 0.792 & 0.732 & 0.221 & 0.045 & 0.653 & 3 & 0 & 0.655 & 3 & 0 & 0.508 & 1 & 0 \\
65 & 1.801 & 0.217 & 3981 & 8985 & 0.387 & 0.224 & 0.405 & 0.033 & 1.527 & 6 & 6 & 1.534 & 5 & 4 & 1.282 & 3 & 0 \\
66 & 1.157 & 0.167 & 1373 & 2851 & 0.366 & 0.305 & 6.335 & 0.008 & 1.099 & 4 & 8 & 1.054 & 4 & 1 & 0.849 & 1 & 0 \\
67 & 1.807 & 0.182 & 4214 & 7354 & 0.826 & 0.228 & 0.342 & 0.036 & 1.209 & 8 & 1 & 1.172 & 5 & 1 & 0.618 & 1 & 0 \\
68 & 0.127 & 0.068 & 2644 & 2811 & 0.953 & 0.874 & 0.098 & 0.067 & 0.000 & 0 & 0 & 0.000 & 0 & 0 & 0.000 & 0 & 0 \\
69 & 4.597 & 0.150 & 17103 & 25626 & 0.703 & 0.423 & 0.013 & 0.184 & 1.322 & 8 & 7 & 1.489 & 7 & 11 & 1.467 & 5 & 2 \\
70 & 0.187 & 0.119 & 1931 & 2431 & 0.787 & 0.637 & 0.367 & 0.035 & 0.000 & 0 & 0 & 0.000 & 0 & 0 & 0.000 & 0 & 0 \\
71 & 0.622 & 0.356 & 1476 & 2322 & 0.699 & 0.368 & 2.744 & 0.013 & 0.341 & 1 & 0 & 0.337 & 1 & 0 & 0.332 & 1 & 0 \\
72 & 0.442 & 0.090 & 4432 & 6119 & 0.761 & 0.499 & 0.072 & 0.078 & 0.000 & 0 & 0 & 0.000 & 0 & 0 & 0.000 & 0 & 0 \\
73 & 2.173 & 0.323 & 6611 & 10977 & 0.745 & 0.293 & 0.107 & 0.064 & 0.868 & 6 & 2 & 0.828 & 4 & 1 & 0.419 & 1 & 0 \\
74 & 0.037 & 0.077 & 1079 & 1418 & 0.684 & 0.644 & 0.413 & 0.033 & 0.000 & 0 & 0 & 0.000 & 0 & 0 & 0.000 & 0 & 0 \\
75 & 0.424 & 0.085 & 4332 & 5173 & 0.841 & 0.699 & 0.074 & 0.077 & 0.000 & 0 & 0 & 0.000 & 0 & 0 & 0.000 & 0 & 0 \\
76 & 0.357 & 0.087 & 1959 & 2463 & 0.826 & 0.609 & 0.673 & 0.026 & 0.227 & 1 & 1 & 0.210 & 1 & 0 & 0.211 & 1 & 0 \\
77 & 0.290 & 0.080 & 967 & 1853 & 0.544 & 0.262 & 4.545 & 0.010 & 0.248 & 1 & 0 & 0.234 & 1 & 0 & 0.237 & 1 & 0 \\
78 & 0.110 & 0.071 & 1073 & 1958 & 0.591 & 0.278 & 1.258 & 0.019 & 0.000 & 0 & 0 & 0.000 & 0 & 0 & 0.000 & 0 & 0 \\
79 & 0.461 & 0.221 & 1214 & 2235 & 0.494 & 0.324 & 3.655 & 0.011 & 0.246 & 0 & 5 & 0.228 & 2 & 0 & 0.179 & 1 & 0 \\
80 & 0.847 & 0.069 & 5071 & 7475 & 0.719 & 0.434 & 0.092 & 0.069 & 0.439 & 1 & 0 & 0.455 & 1 & 1 & 0.451 & 1 & 0 \\
81 & 1.815 & 0.101 & 6481 & 8012 & 0.877 & 0.603 & 0.095 & 0.068 & 1.509 & 5 & 3 & 1.266 & 5 & 3 & 1.309 & 5 & 0 \\
82 & 0.664 & 0.097 & 1077 & 1258 & 0.990 & 0.634 & 7.537 & 0.008 & 0.634 & 1 & 0 & 0.633 & 1 & 0 & 0.629 & 1 & 0 \\
83 & 1.332 & 0.100 & 9262 & 15025 & 0.607 & 0.386 & 0.024 & 0.136 & 0.296 & 1 & 4 & 0.312 & 2 & 3 & 0.145 & 1 & 0 \\
84 & 0.994 & 0.214 & 3394 & 5859 & 0.977 & 0.199 & 0.361 & 0.035 & 0.000 & 0 & 0 & 0.000 & 0 & 0 & 0.000 & 0 & 0 \\
85 & 0.337 & 0.048 & 2805 & 4211 & 0.691 & 0.428 & 0.216 & 0.045 & 0.000 & 0 & 0 & 0.000 & 0 & 0 & 0.000 & 0 & 0 \\
86 & 0.410 & 0.099 & 1980 & 2052 & 0.969 & 0.926 & 0.750 & 0.024 & 0.322 & 1 & 4 & 0.284 & 2 & 0 & 0.260 & 1 & 0 \\
87 & 0.182 & 0.067 & 2042 & 5611 & 0.277 & 0.174 & 0.303 & 0.038 & 0.000 & 0 & 0 & 0.000 & 0 & 0 & 0.000 & 0 & 0 \\
88 & 0.368 & 0.134 & 2551 & 3441 & 0.755 & 0.539 & 0.314 & 0.038 & 0.000 & 0 & 0 & 0.000 & 0 & 0 & 0.000 & 0 & 0 \\
89 & 0.383 & 0.074 & 3464 & 4575 & 0.749 & 0.580 & 0.131 & 0.058 & 0.000 & 0 & 0 & 0.000 & 0 & 0 & 0.000 & 0 & 0 \\
90 & 1.273 & 0.113 & 3135 & 4948 & 0.557 & 0.456 & 0.586 & 0.027 & 1.059 & 5 & 2 & 1.188 & 5 & 2 & 0.825 & 1 & 0 \\
91 & 1.208 & 0.064 & 8357 & 18062 & 0.393 & 0.252 & 0.029 & 0.123 & 0.000 & 0 & 0 & 0.000 & 0 & 0 & 0.000 & 0 & 0 \\
92 & 0.647 & 0.115 & 4163 & 5090 & 0.857 & 0.639 & 0.127 & 0.059 & 0.000 & 0 & 0 & 0.000 & 0 & 0 & 0.000 & 0 & 0 \\
93 & 2.737 & 0.066 & 9102 & 10841 & 0.847 & 0.699 & 0.051 & 0.093 & 2.427 & 6 & 6 & 2.402 & 4 & 6 & 1.662 & 1 & 0 \\
94 & 0.359 & 0.276 & 1384 & 2865 & 0.389 & 0.290 & 1.923 & 0.015 & 0.000 & 0 & 0 & 0.000 & 0 & 0 & 0.000 & 0 & 0 \\
95 & 0.365 & 0.094 & 1262 & 1830 & 0.573 & 0.573 & 2.573 & 0.013 & 0.329 & 1 & 4 & 0.274 & 1 & 0 & 0.282 & 1 & 0 \\
96 & 0.644 & 0.108 & 3642 & 5406 & 0.575 & 0.531 & 0.189 & 0.048 & 0.466 & 1 & 5 & 0.363 & 1 & 0 & 0.378 & 1 & 0 \\
97 & 1.055 & 0.074 & 6965 & 10928 & 0.562 & 0.461 & 0.044 & 0.100 & 0.521 & 3 & 3 & 0.540 & 3 & 0 & 0.449 & 1 & 0 \\
98 & 3.319 & 0.117 & 4889 & 8811 & 0.462 & 0.370 & 0.403 & 0.033 & 2.606 & 8 & 34 & 3.142 & 8 & 2 & 2.977 & 5 & 0 \\
99 & 0.691 & 0.197 & 4088 & 8322 & 0.492 & 0.241 & 0.144 & 0.056 & 0.000 & 0 & 0 & 0.000 & 0 & 0 & 0.000 & 0 & 0 \\
100 & 0.093 & 0.059 & 889 & 1324 & 0.640 & 0.474 & 1.885 & 0.015 & 0.000 & 0 & 0 & 0.000 & 0 & 0 & 0.000 & 0 & 0 \\
\hline
&&&&&&&&&&&&&&&&&\\
Total & 99.7\footnote{ The total mass of all cores. The total mass of only the prestellar cores is $75.7\,\mathrm{M}_\odot$} &&&&&&&& 55.5 & 200 & 217 & 54.1 & 186 & 89 & 44.6 & 93 & 5

\label{onlinematerial}
\end{longtable}

\end{landscape}

\label{lastpage}
\end{document}